\documentclass[pra,onecolumn,showpacs,showkeys]{revtex4}
\usepackage{natbib}
\usepackage{graphics}
\usepackage{amsmath}
\usepackage{mathrsfs}
\usepackage{theorem}
\usepackage{overpic}
\usepackage{hyperref}
\usepackage{rotating}
\usepackage{amssymb}
\usepackage[english]{babel}
\usepackage{color,times}
\usepackage{CJK}

\newcommand{\ket}[1]{|#1\rangle}
\newcommand{\bra}[1]{\langle #1|}

\newcommand{\be}{\begin{eqnarray}}
\newcommand{\ee}{\end{eqnarray}}

\begin{document}
\begin{CJK*}{GB}{gbsn}
\title{Detecting ground-state degeneracy in many-body systems through qubit decoherence}
\author{H. T. Cui(´Þº£ÌÎ)$^ {a,b}$}
\email{cuiht@aynu.edu.cn}
\author{X. X. Yi(ÒÂѧϲ)$^{b}$}
\email{yixx@nenu.edu.cn}
\affiliation{$^a$School of Physics and Electric Engineering, Anyang Normal University, Anyang 455000, China}
\affiliation{$^b$Center for Quantum Sciences, Northeast Normal University, Changchun 130024, China}
\date{\today}

\begin{abstract}
By coupling with a qubit, we demonstrate that qubit decoherence can unambiguously detect the occurrence of ground-state degeneracy in many-body systems. We first demonstrate universality using the two-band model. Consequently, several exemplifications, focused on topological condensed matter systems in one, two, and three dimensions, are presented to validate our proposal. The key point is that qubit decoherence varies significantly when energy bands touch each other at the Fermi surface. In addition, it can partially reflect the degeneracy inside the band. This feature implies that qubit decoherence can be used for reliable diagnosis of ground-state degeneracy.
\end{abstract}
\pacs{03.65.Yz, 03.65.Vf, 73.20.At, 73.43.-f }
\keywords{decoherence, quantum phase transition, ground-state degeneracy}
\maketitle
\end{CJK*}

\section{Introduction}
Ground-state degeneracy is closely related to the exotic features in many-body systems. Particularly, topological order and related boundary modes are observed when ground-state degeneracy occurs \cite{tknn82, wen}. Thus, detecting the occurrence of ground-state degeneracy is an interesting issue. One can directly calculate the energy spectrum in an exact manner. However, this is not possible for realistic situations. Numerical or first-principle calculation of the energy band is another typical method. However, the obtained results are often ambiguous and indecisive because of computational precision. In experiments, high resolution angle-resolved photoemission spectroscopy is extensively used in real materials. However, these materials impose a considerable requirement on the experiment\cite{dhs}. In addition, this method is insensitive to bulk properties in systems. Thus, finding a simple and efficient method to detect ground state degeneracy is important.

 Quan \emph{et al.,} showed that the qubit coherence, characterized by the Loschmidt echo, decays rapidly when the spin-chain environment exhibits phase transition \cite{quan}. This implies the possibility of indirect detection of ground-state degeneracy in many-body systems. In this study, we first demonstrate that this feature is universal, using the two-band theory. Then, through exemplifications, we demonstrate that qubit decoherence can be used for reliable detection of ground-state degeneracy. This finding is interesting for theoretical or experimental consideration as a qubit is weakly coupled with the system that we studied. Moreover, this study is also interesting to the recent topic of preserving spin coherence\cite{yang}.

\section{Two-band theory}
 It is natural to use the two-band model as the starting point, which can be written as
\be\label{h}
H = \sum_{x, x'}\mathbf{c}_x^{\dagger}\mathscr{H}_{x, x'}\mathbf{c}_{x'},
\ee
where $\mathbf{c}_x^{(\dagger)}= (c_+, c_-)_x^{\textrm{T}(\dagger)}$, specified by concrete Hamiltonians, defines a pair of fermion annihilation (creation) operators at site $x$. $\mathscr{H}_{x, x'}$ is a $2\times2$ matrix, which can be written as
\be\label{hx}
\mathscr{H}_{x,x'}=\left(\begin{array}{cc}t_+ & \lambda \\ \lambda' & t_-
\end{array}\right)_{x,x'}
\ee
where $t_{\pm, xx'}=t_{\pm,x'x}^*$ and $\lambda_{x,x'}=\lambda^*_{x',x}$ with the requirement of hermiticity. Despite being simple, Hamiltonian Eq. \eqref{h} has a wide range of applications, such as the Bogoliubov-de Gennes Hamiltonian in superconductivity, and graphite systems \cite{ryu}.

Without loss of generality, it is conventional to impose a periodic boundary condition. Then using the Fourier transformation, $\mathbf{c}_x=1/\sqrt{N}\sum_k e^{ikx}\mathbf{c}_k$,  where $k=2\pi n/N$ with $n=1,2,\dots, N$,  Eq. \eqref{h} can be written as $H = \sum_k \mathbf{c}_k^{\dagger}\mathscr{H}(k)\mathbf{c}_k$. Introducing a four-vector, $R_{\mu}(k)(\mu=0,x,y,z)$, we obtain
\be\label{hk}
\mathscr{H}(k)=\sum_{\mu}R_{\mu}(k)\sigma_{\mu},
\ee
where $k$ spans the first Brillouin zone, $\sigma_0$ denotes a $2\times 2$ unit matrix, and $\sigma_{\alpha}(\alpha=x,y,z)$ represents the Pauli matrices. Eq. \eqref{hk} can be diagonalized using the eigenvectors, $\nu_{\pm}$, of $\mathbf{R}(k)\cdot\mathbf{\sigma}$,
\be\label{nu}
\nu_{\pm}= \frac{1}{\sqrt{2R(k)(R(k)\mp
R_{z}(k))}}\left(\begin{array}{c}R_{x}(k) - i R_{y}(k)\\ \pm
R(k)-R_{z}(k)\end{array}\right)
\ee
where $R(k)=|\mathbf{R}(k)|$, and eigenvalues  $E_{\pm}(k)=R_0(k) \pm R(k)$. The ground state is defined as the filled Fermi sea,  $\ket{g}=\prod_{E_{-}(k)} \beta_{-, k}^{\dagger}\ket{0}_k$, where $\beta_{-, k}^{\dagger}=\mathbf{c}_k^{\dagger}\nu_- $ and $\ket{0}_k$ is the vacuum state. Clearly, there is an energy gap, $2R(k)$, between two bands, $E_+ (k)$ and $E_- (k)$. Ground-state degeneracy occurs when $R(k)=0$. For topologically nontrivial systems, energy level crossing induces boundary states, which exhibit nontrivial topological features, e.g., chirality or dissipationless transport \cite{ti}.

To detect the occurrence of degeneracy, a qubit is introduced,
\be \label{qubit}
H_{\text{int}}=\delta \ket{1}\bra{1} \otimes \sum_x \mathbf{c}_x^{\dagger} \sigma_z \mathbf{c}_x,
\ee
where $\ket{1}$ denotes the higher energy level of the qubit \cite{quan}.  We suppose that the qubit interacts differently with each primitive unit cell in a lattice. Thus, $\delta$ denotes the difference between couplings, which is significantly weaker than typical couplings in $H$. Eq. \eqref{qubit} corresponds to dephasing of the qubit, and is a clear choice as it contributes only to the diagonal elements in $\mathscr{H}(k)$, and thus, cannot change the physics in $H$. Then, the total Hamiltonian is
\be
\mathcal{H}=H+H_{\text{int}}.
\ee

Suppose the initial state is $\ket{\psi(t=0)}=(c_0 \ket{0} + c_1 \ket{1})\otimes\ket{g}$, then
\be
\ket{\psi(t)}=e^{- i\mathcal{H}t}(c_0 \ket{0} + c_1 \ket{1})\otimes\ket{g}=(c_0 \ket{0} e^{- iHt} + c_1 \ket{1}e^{- i\mathcal{H}t} ) \otimes\ket{g}\nonumber.
\ee
Then, up to a trivial phase factor for a non-diagonal element, the decoherence factor can be defined as
\be
\mathcal{L}(t)=\bra{g}e^{- i\mathcal{H}t}\ket{g}.
\ee
The eigenstates and eigenvalues of $\mathcal{H}$ can be obtained by replacing $R_z(k)$ with $\delta + R_z(k)$ in Eq. \eqref{hk}. Then
\be \label{df}
\mathcal{L}(t)&=&\prod_k e^{- iR_0(k)t} \cdot \mathcal{L}_k(t)
\ee
where
\be\label{dfk}
 \mathcal{L}_k(t)&=&\cos[R(k, \delta)t] + i \tfrac{R(k)^2 - \delta R_z(k)}{R(k) R(k,\delta)} \sin[R(k, \delta)t], \nonumber \\
R(k,\delta)&=&\sqrt{R_x(k)^2 + R_y(k)^2 + \left[(R_z(k)+\delta\right]^2};\nonumber
\ee
It is clear that a singularity exists when $R(k)=0$, for which the energy gap vanishes. Then, modulus $\mathcal{L}(t)$ and $ \mathcal{L}_k(t)$  change abruptly.

\section{Exemplifications}

The \emph{Su-Schrieffer-Heeger (SSH) model} depicts a one-dimensional system, which can exhibit topological features. The tight-binding Hamiltonian is \cite{heeger}
\be\label{ssh}
H=\sum_{l=1}^{N}(-1 + (-1)^l\phi_l)(c_l^{\dagger}c_{l+1}+ h.c.),
\ee
where $\phi_l$ represents the dimerization at the $l$-th site, and the alternating sign of the hopping elements represents the dimerization between carbon atoms in a molecule. Without loss of generality, it is convenient to neglect the kinetic energy of the system and consider $\phi_l=\phi$ \cite{ryu2}. Using a periodic boundary condition, the SSH model can be written as \cite{ryu2}
\be
\mathscr{H}(k)=\left(\begin{array}{cc} 0 & -(1 + \phi)- (1 - \phi) e^{- ik} \\ -(1 + \phi)- (1 - \phi) e^{ ik} & 0 \end{array}\right).
\ee
where $k\in[-\pi, \pi]$. Then, $R(k)=2\sqrt{\cos^2 \tfrac{k}{2}+\phi^2 \sin^2 \tfrac{k}{2}}$.

It is clear that there exists a band crossing at boundary $k=\pi$ when $\phi=0$. Furthermore, even though the Chern number cannot be defined for one-dimensional systems, the ground states can be identified as two distinct phases by introducing the quantized Berry phase, which is $\pi$ for $\phi\in[-1, 0)$ and 0 for $\phi\in(0, 1]$ \cite{ryu2}.

\begin{figure}[tbp]
\center
\includegraphics[width=7cm]{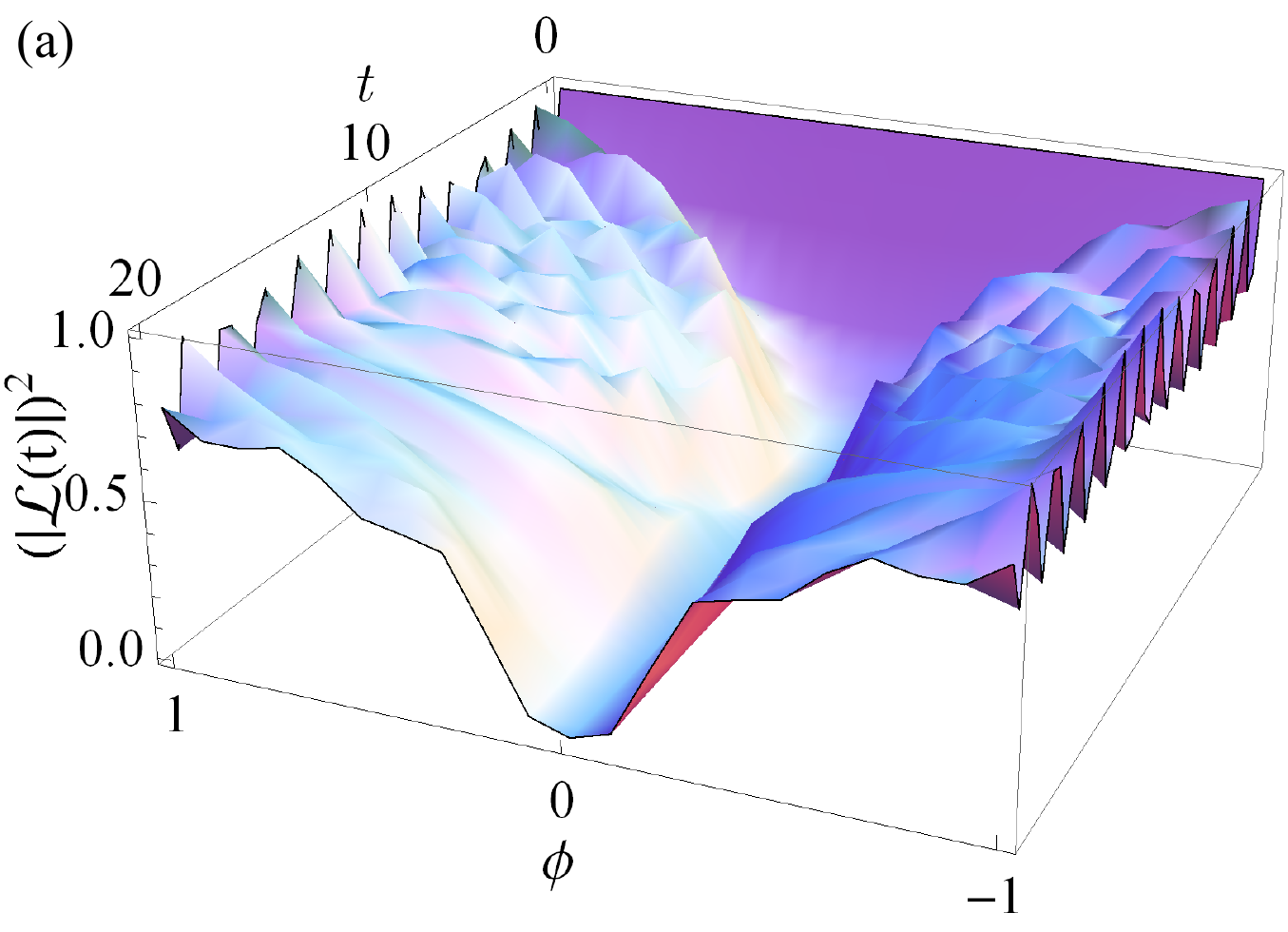}
\includegraphics[width=7cm]{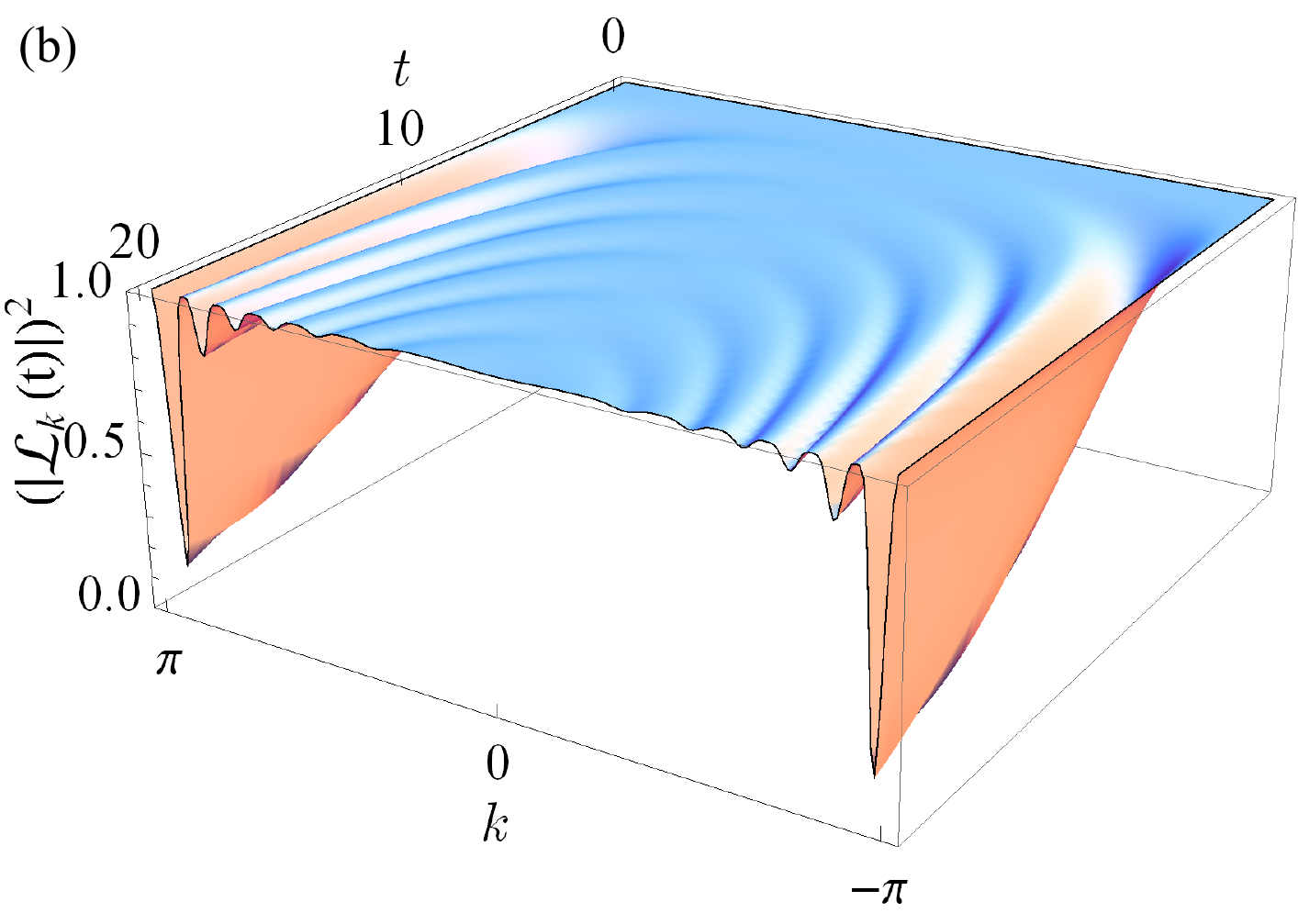}
\caption{(Color online) (a) $|\mathcal{L}(t)|^2$  and (b) $|\mathcal{L}_k(t)|^2$ for SSH model when $N=200$ and  $\delta=0.1$.  For (b) we chose $\phi=0$.}
\label{fig:SSH}
\end{figure}

By coupling with a qubit, we obtain
\be
\mathcal{L}(t)=\prod_k \mathcal{L}_k(t)=\prod_k \left\{\cos[R(k, \delta)t] + i \tfrac{R(k)}{R(k, \delta)} \sin[R(k, \delta)t]\right\},
\ee
where
\be
R(k, \delta)&=&2\sqrt{\cos^2 \tfrac{k}{2}+\phi^2 \sin^2 \tfrac{k}{2}+ \tfrac{\delta^2}{4}}. \nonumber
\ee
As shown in Fig. \ref{fig:SSH}, the modulus of $\mathcal{L}(t)$ exhibits a cusp at the critical point, $\phi=0$. $|\mathcal{L}_k(t)|^2$ exhibits strong decay close to $k=\pm \pi$, which is a characteristic of the occurrence of degeneracy.

\begin{figure}[tbp]
\center
\includegraphics[width=7cm]{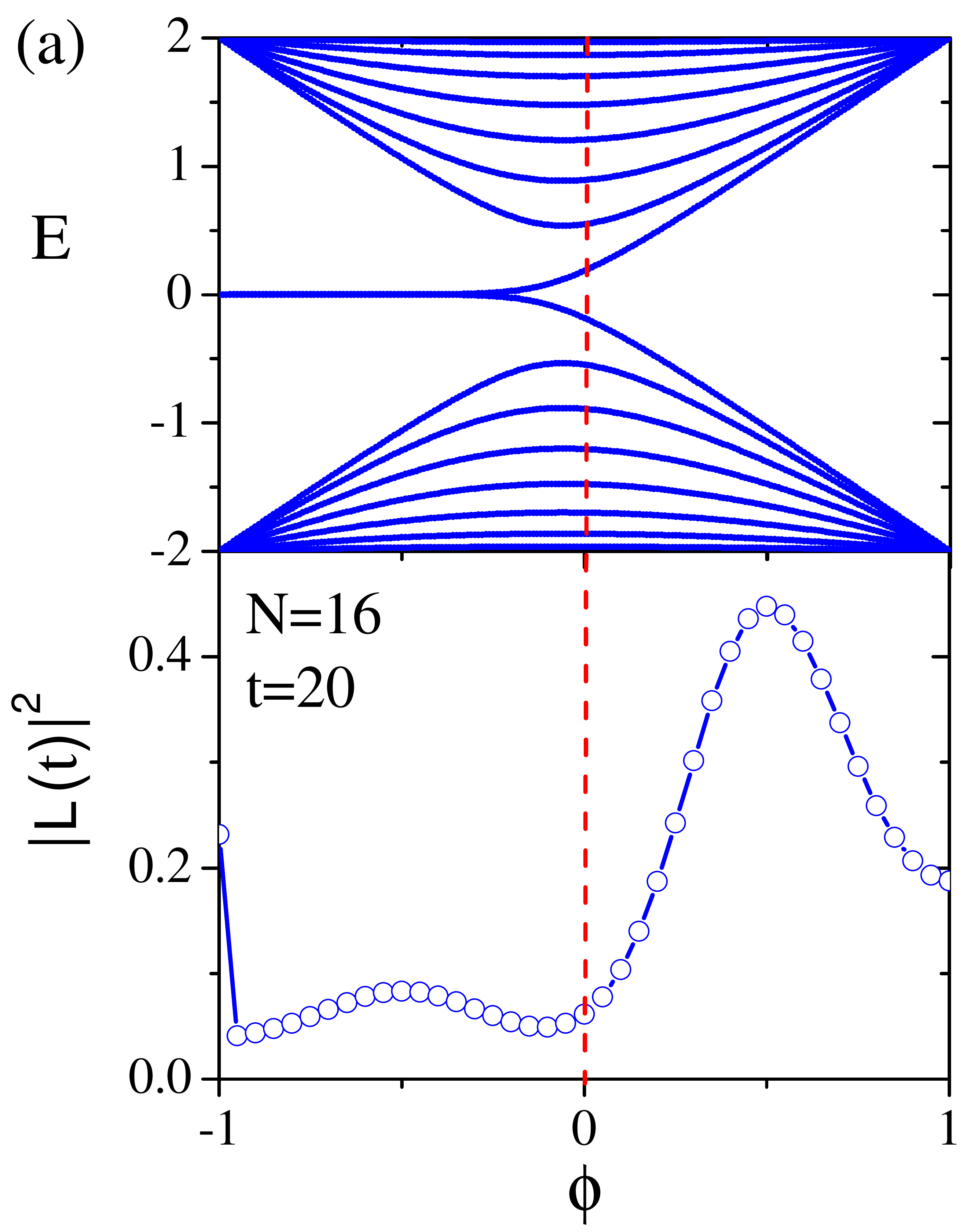}
\includegraphics[width=7cm]{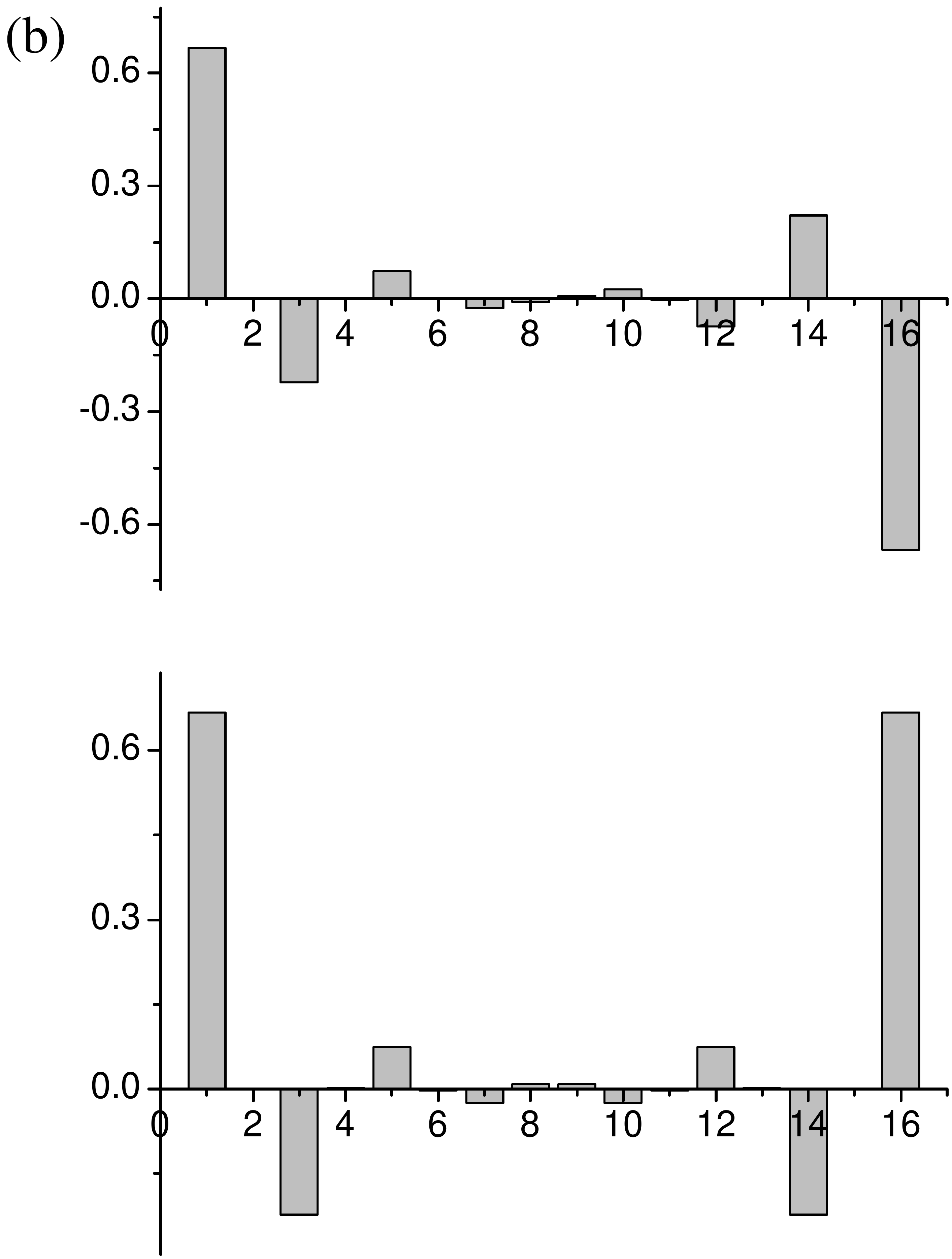}
\caption{(Color online) (a) $|\mathcal{L}(t)|^2$  for open-boundary SSH model when $\delta=0.1$.  For (b) the two boundary states are presented for $\phi=0.5$.}
\label{fig:SSHopen}
\end{figure}

For an open boundary, the SSH model exhibits two degenerate boundary states, as shown in Fig. \ref{fig:SSHopen}(b), which originate from the topology in the model. $|\mathcal{L}(t)|^2$ and the energy spectrum are shown in Fig. \ref{fig:SSHopen}(a), obtained using exact calculation. Owing to limitations in computing resource, a finite lattice number is considered. Even though the phase transition point deviates from $\phi=0$ because of the finite lattice number, $|\mathcal{L}(t)|^2$ represents different features in the regions $\phi<0$ and $\phi>0$.

\emph{Qi--Wu--Zhang (QWZ) model}. Using a periodic boundary condition, the Hamiltonian can be written as \cite{QWZ}
\be \label{qwz}
\mathscr{H}(k)=\sigma_x \sin k_x + \sigma_y  \sin k_y + \sigma_z (2 + M - \cos k_x - \cos k_y),
\ee
which characterizes a topological insulator in a two-dimensional system. Eq. \eqref{qwz} is fully gapped except at the following critical values of $M$: at $M=0$ for $(k_x, k_y)= (0, 0)$; at $M=-2$ for $(k_x, k_y)= (\pi, 0), (0, \pi)$; and at $M=-4$ for $(k_x, k_y)= (\pi, \pi)$. The degeneracy points separate three distinct regions represented by the following Chern numbers: 0 for $M>0$ and $M<-4$; 1 for $-4<M<-2$; and -1 for  $-2<M<0$. The Chern number can be changed only by closing the energy gap, for which the topological edge states, which connect lower and upper bands, can occur at the boundary.

\begin{figure}[t]
\center
\includegraphics[width=7.5cm]{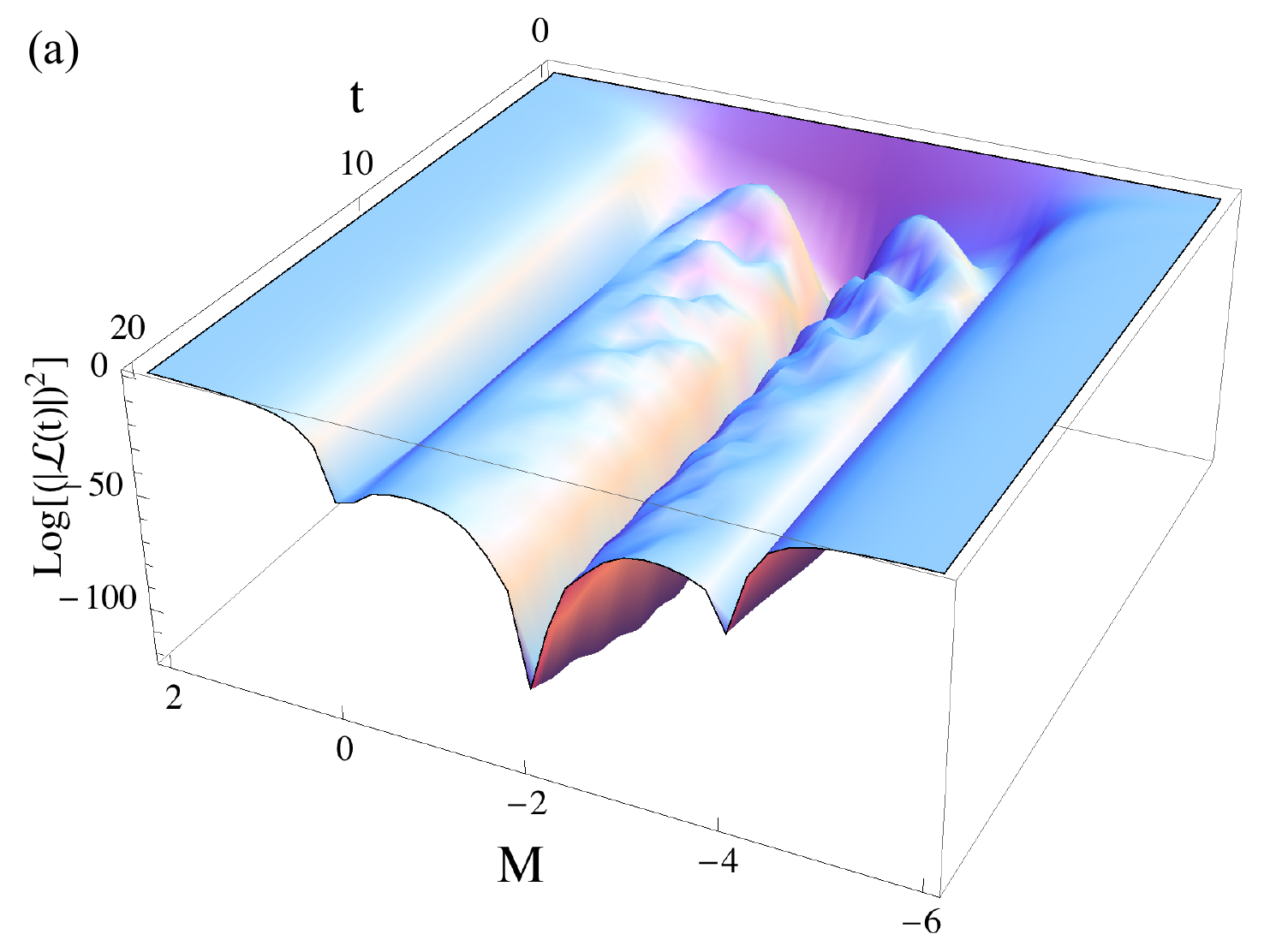}
\includegraphics[width=7.5cm]{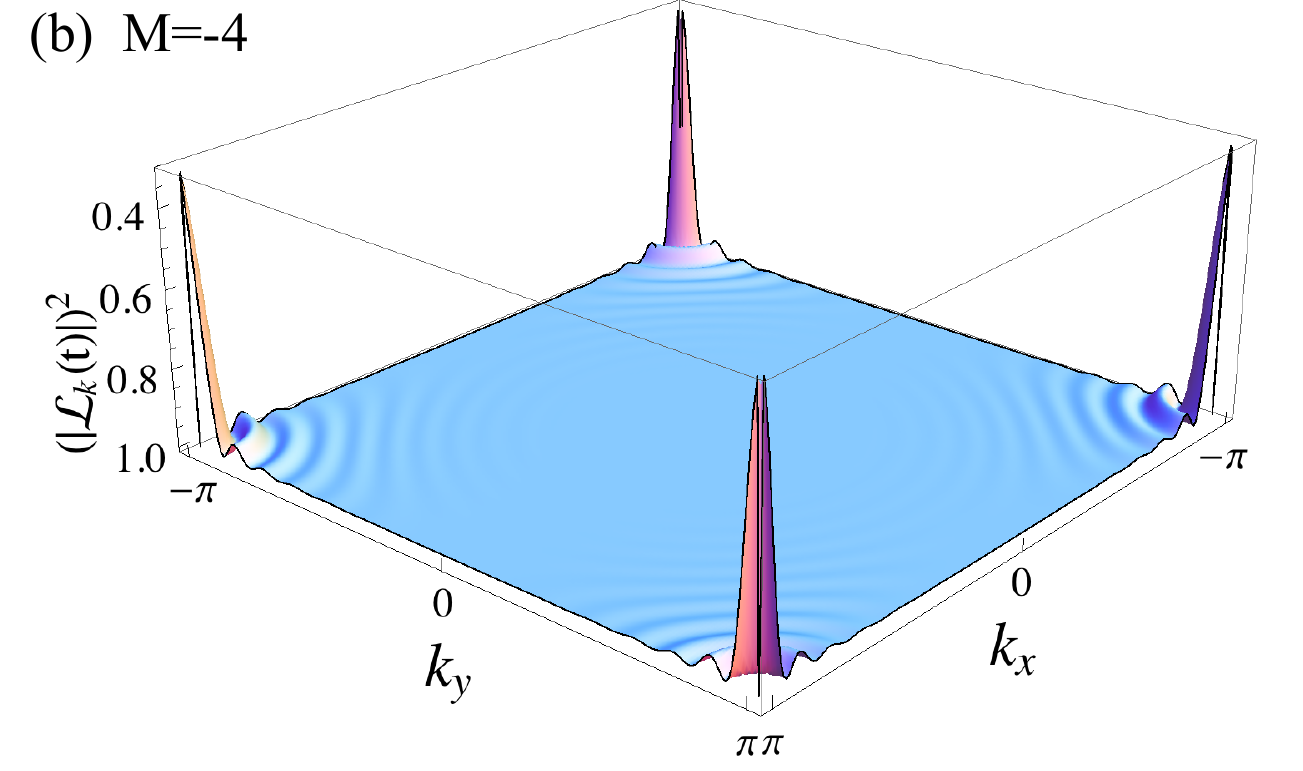}
\includegraphics[width=7.5cm]{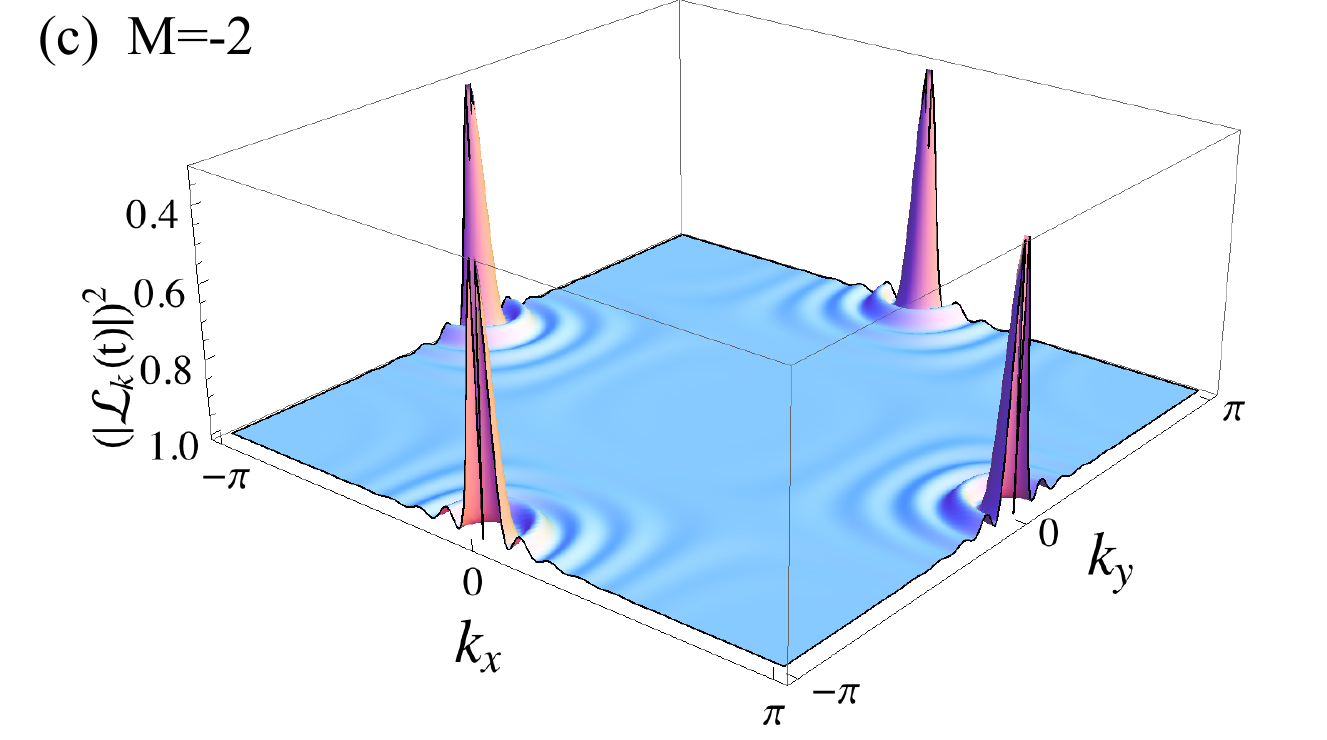}
\includegraphics[width=7.5cm]{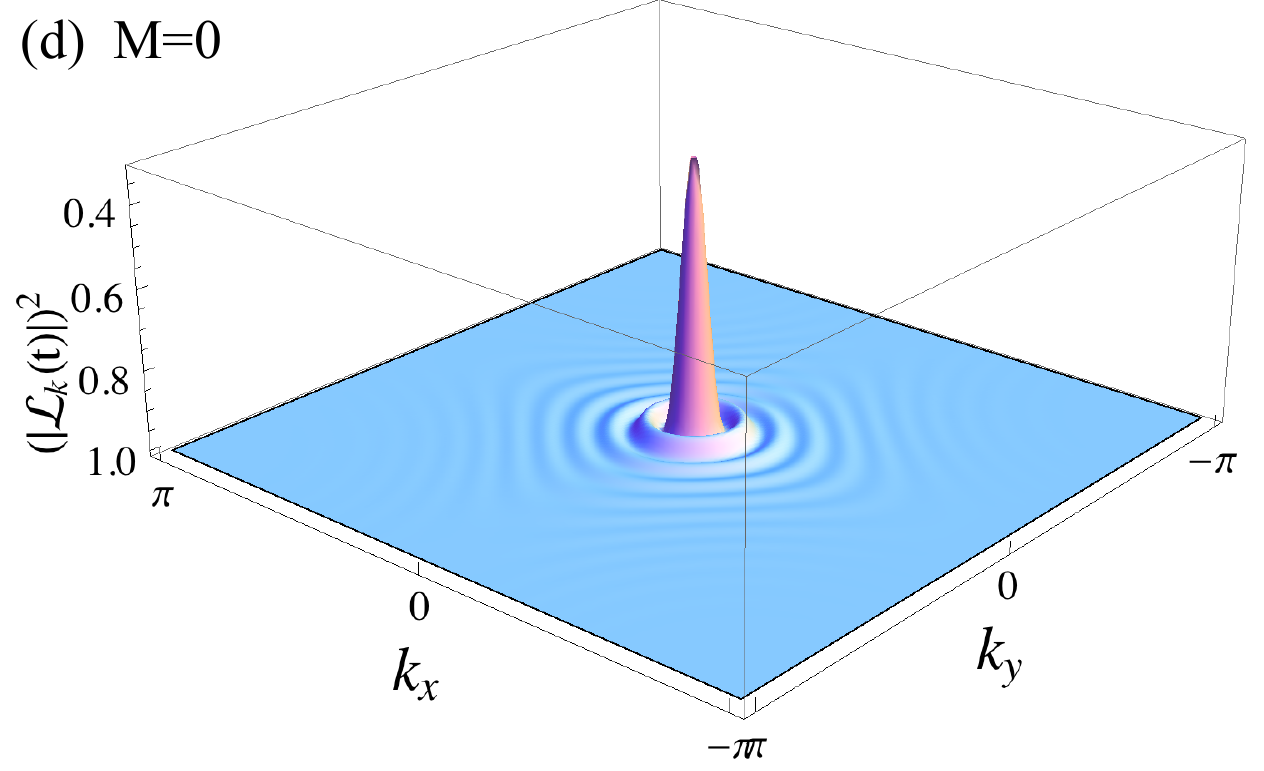}
\caption{(Color online) The logarithm of  $|\mathcal{L}(t)|^2$   and  $|\mathcal{L}_k(t)|^2$ for different  $M$ in QWZ model when $N_x=N_y=100$ and $\delta=0.1$. For plottings of (b)-(d), $t=20$ is chosen and the lattice constant is set to be 1. }
\label{fig:qwz}
\end{figure}

By coupling with a qubit, we obtain
\be \small
R(k, \lambda)=\sqrt{\sin^2 k_x + \sin^2 k_y+ (2 + M + \delta - \cos k_x - \cos k_y)^2}. \nonumber
\ee
A plot of the modulus of $\mathcal{L}(t)$ is presented in Fig. \ref{fig:qwz}, in which there are three cusps, which occur exactly at the critical points. It should be noted that we use the logarithm of $|\mathcal{L}(t)|^2$ with base 10 in this graph as $|\mathcal{L}(t)|^2$ tends toward zero rapidly for large $N_x \times N_y$. The corresponding $(k_x, k_y)$ can be observed clearly by plotting $|\mathcal{L}_k(t)|^2$, as shown in Figs. \ref{fig:qwz} (b), (c), and (d). It is clear that $|\mathcal{L}_k(t)|^2$ changes abruptly when approaching the degeneracy points.

\begin{figure}[t]
\center
\includegraphics[width=7cm]{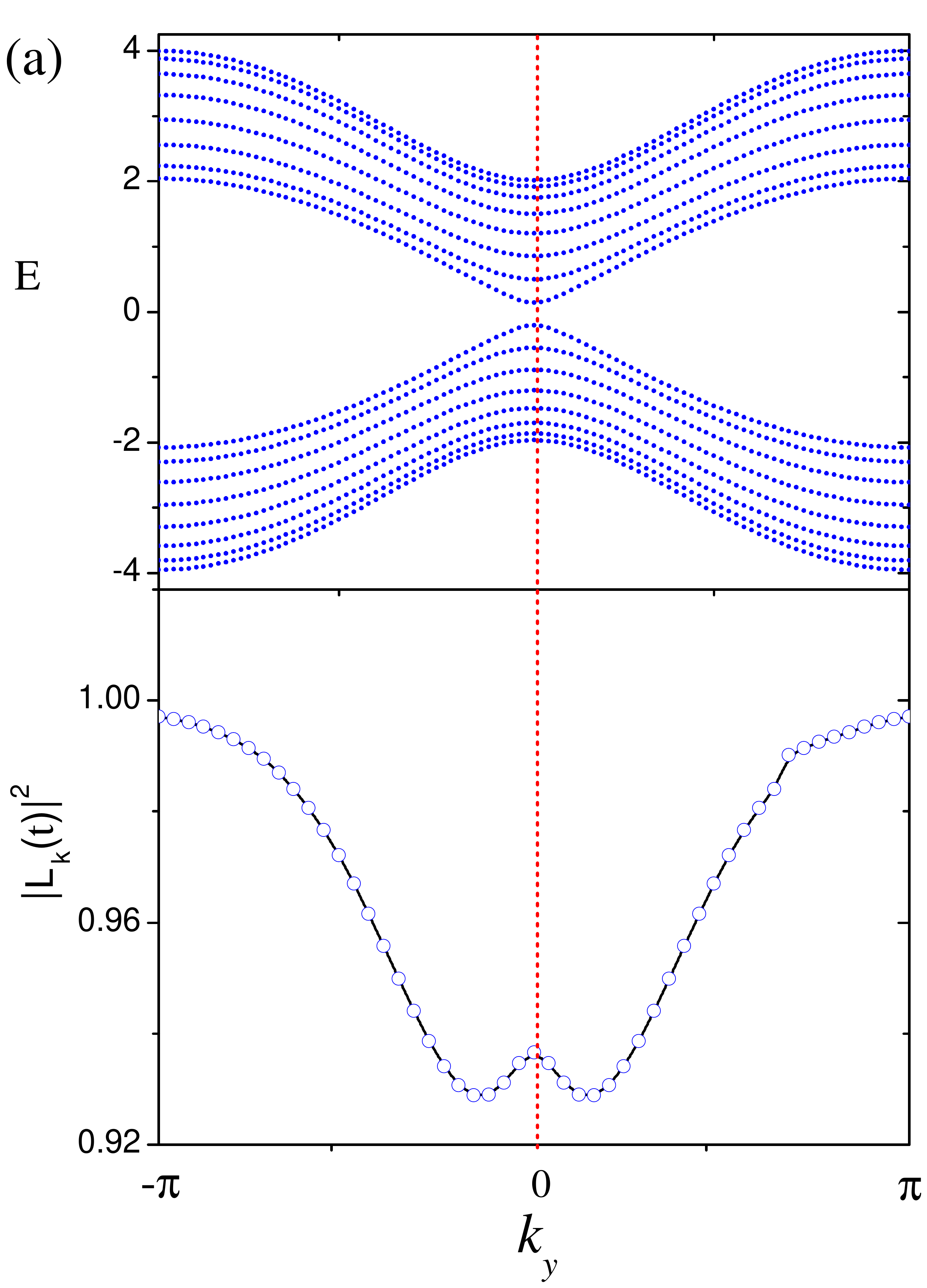}
\includegraphics[width=7cm]{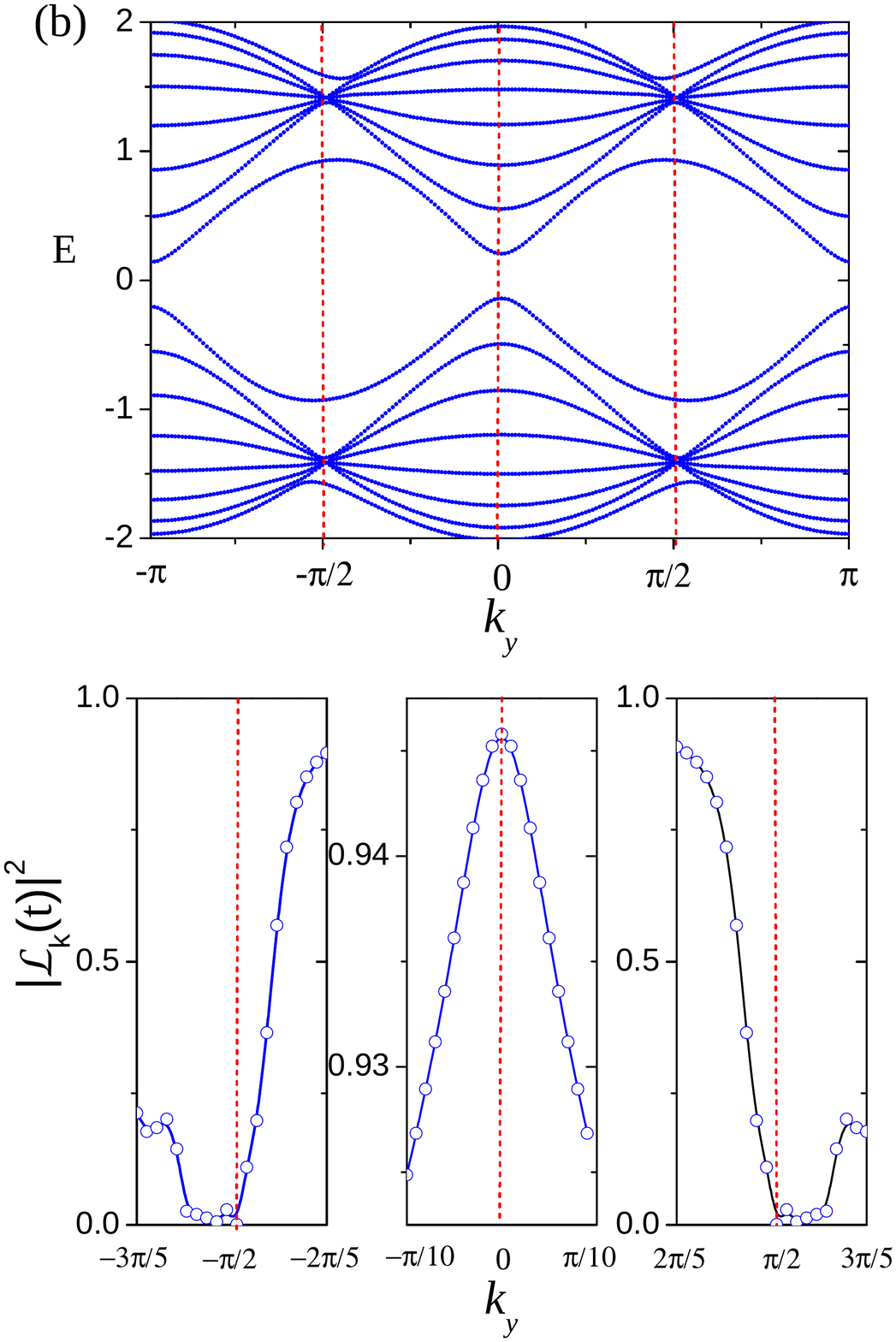}
\caption{(Color online) The energy spectrum and  $|\mathcal{L}_k(t)|^2$  for open-boundary QWZ model when $N=16$ and $\delta=0.1$, which are obtained by exact evaluation.  We have chosen that the systems is open at $x$ direction.  For panel (a) $M=0$. As for panels (b),  $M=-2$ and  only three special regions of $k_y a$ are chosen to plot. The situation for $M=-4$ is similar to that of $M=0$, and thus does not be shown in this place.}
\label{fig:qwzopen}
\end{figure}

\begin{figure}[t]
\center
\includegraphics[width=8cm]{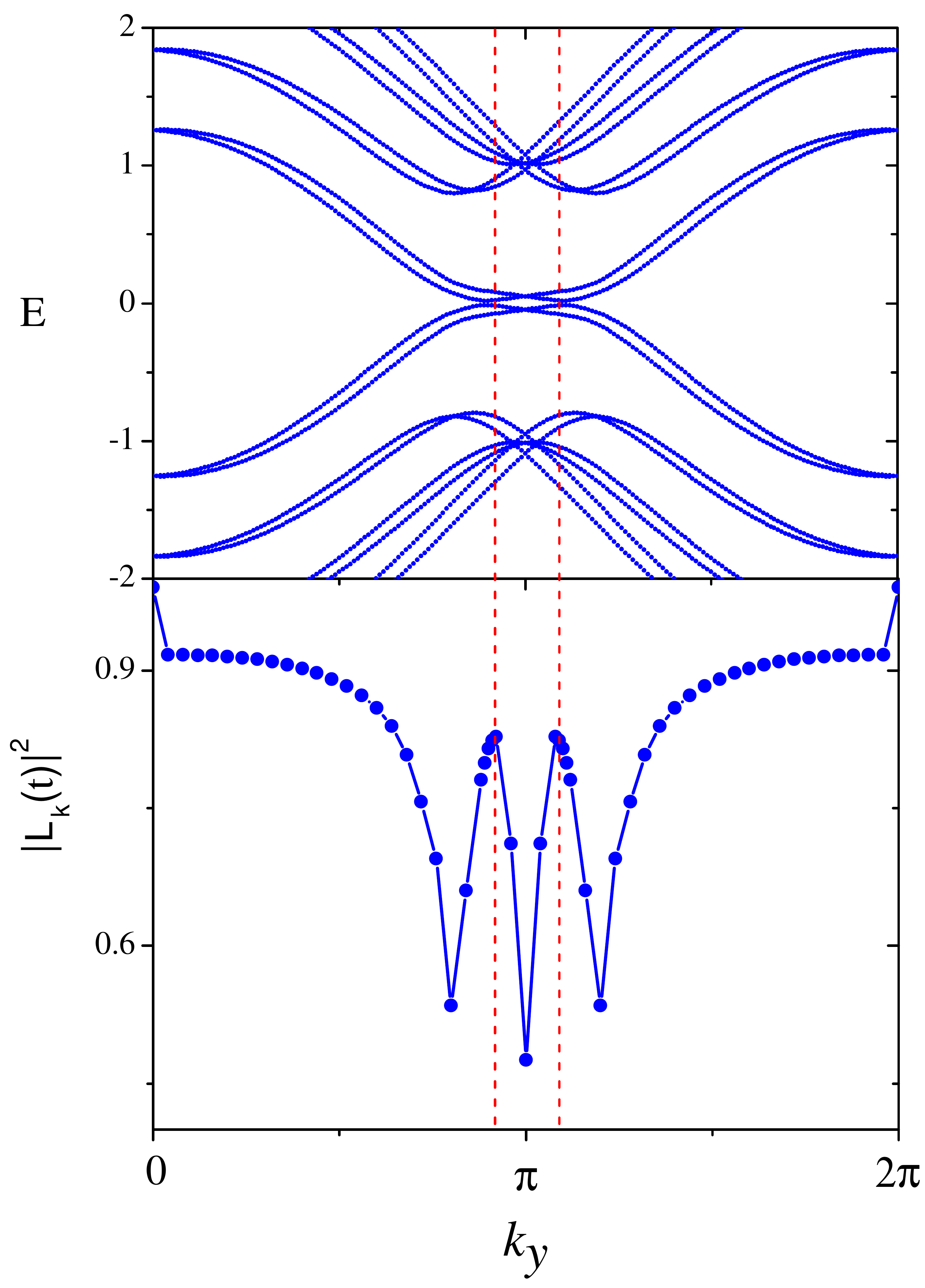}
\caption{(Color online) A plotting of $|\mathcal{L}_k(t)|^2$ for KM model when $\delta=0.1$ and $t=20$, for which $\lambda_{SO}=0.06$, $\lambda_R=0.05 $ and $\lambda_v=0.1 /\sqrt{3}$. For the upper plotting, $N_A=N_B=100$ at $y$ direction and $N_A=N_B=8$ at $x$ direction. For the lower plotting, $N_A=N_B=25$ at $y$ direction and $N_A=N_B=8$ at $x$ direction are chosen in this exact evaluation.  }
\label{fig:km}
\end{figure}

Chiral edge states can occur at an open boundary \cite{QWZ}. At the phase transition points, the energy gap displays avoided crossings at certain values of $k_y$ because of a finite number of lattice points, as shown in Fig. \ref{fig:qwzopen}. $|\mathcal{L}_k(t)|^2$ varies significantly when the energy gap tends to close. In addition, it should be noted that the degeneracy inside the band reduces $|\mathcal{L}_k(t)|^2$, as shown for $k_y a= \pm \pi/2$ in Fig. \ref{fig:qwzopen}(c). This feature implies that qubit decoherence can partially reflect the property inside the band. A similar phenomenon is observed in the following studies:

\emph{Kane--Mele (KM) model}. A more interesting case is incorporating spin-orbit interaction in the model. Kane and Mele studied this issue first, and found that there exist edge states connecting the valence and conduction bands without breaking time-reversal invariance, which is currently well known as the quantum spin Hall effect (QSHE) \cite{km}. Consider the following Hamiltonian of graphene \cite{km}:
\be \label{km}
H=\sum_{\langle i, j \rangle, \alpha } c^\dagger_{i, \alpha}c_{j, \alpha} + \sum_{\langle\langle i, j \rangle\rangle, \alpha, \beta } i \lambda_{SO} v_{ij} s^z_{\alpha\beta} c^\dagger_{i, \alpha}c_{j, \beta} +  i \lambda_R \sum_ {\langle i, j \rangle, \alpha } c_i \left( \mathbf{s} \times  \mathbf{d}_{ij} \right)_z c_j + \lambda_v \sum_i \xi_i c_i^\dag c_i,
\ee
where $\langle i, j \rangle$ and $\langle\langle i, j \rangle\rangle$ denote the nearest and next-nearest neighbor sites, respectively, $s^z$ is the electron spin, $v_{ij}=- v_{ji}=\pm 1$, depending on whether the electron turns left $(+1)$ or right $(-1)$ when traveling from site $j$ to $i$, and $ \mathbf{d}_{ij}$ is a vector that connects sites $i$ and $j$. With a zig-zag boundary in the $x$ direction, two band-touching points are observed at the Fermi surface adjacent to the point $k_y a=\pi$, where $a$ is the lattice constant, as shown in the top panel in Fig. \ref{fig:km}.

By coupling with a qubit, we obtain
\be
H_{int}=\delta \ket{1}\bra{1}\sum_{i, \alpha}c^{\dagger}_{2i-1, \alpha} c_{2i-1, \alpha} - c^{\dagger}_{2i, \alpha} c_{2i, \alpha},
\ee
where the odd and even sites correspond to two primitive unit cells, $A$ and $B$, respectively, in graphene. As shown in Fig. \ref{fig:km}, $|\mathcal{L}_k(t)|^2$ increase abruptly at the degenerate points,  where the energy levels are crossing through Fermi surface $E=0$. Moreover, it decreases rapidly at $k_y a=\pi$, at which no level crossing occurs at the Fermi surface; however, there is degeneracy in the band. A similar feature is observed in the QWZ model, as shown in Fig. \ref{fig:qwzopen}(c). This implies that $|\mathcal{L}_k(t)|^2$ is sensitive to the degeneracy in the band.

\emph{Weyl semimetal model}. Generalizing the QSHE for a three-dimensional case, a new topological phase can occur when either inversion or time-reversal symmetry is broken, which is referred to as Weyl semimetal phase. A typical feature of a Weyl semimetal is that the degeneracy points at the Fermi surface occur in pairs. Thus, a surface state can be introduced instead of an edge state in a two-dimensional case, which exhibits distinct chirality across band-touching points \cite{weyl}.

For simplicity, we focus on a tight-binding model introduced in \cite{vf}, which supports a pair of Weyl nodes. The Hamiltonian is
\be\label{weyl}
\mathscr{H}(k)&=& \mathscr{H}_0(k) + \mathscr{H}_I (k) \nonumber \\
\mathscr{H}_0(k)&=& 2 \lambda \sigma_z (s_x \sin k_y - s_y \sin k_x)+ 2 \lambda_z \sigma_y \sin k_z + \sigma_x M_k \nonumber \\
\mathscr{H}_I(k)&=& b_0 \sigma_y s_y - b_1 \sigma_x s_x + b_2 \sigma_x s_y + b_3 s_z,
\ee
where $M_k= \varepsilon - 2 t\sum_{\alpha=x, y, z} \cos k_{\alpha}$,  and $\sigma_{\alpha}$ and $s_{\alpha}$ characterize the orbital and spin dimension, respectively. $\mathscr{H}_0(k)$ represents a three-dimensional topological insulator, regularized in a simple cubic lattice. In $\mathscr{H}_I (k)$, the term including $b_0$ breaks inversion symmetry and preserves time-reversal symmetry, while the remaining terms have the opposite effect. Hence, because of $\mathscr{H}_I (k)$, the degenerate band-touching points can be separated in moment or energy dimensions, as shown in Figs. \ref{fig:weyl-en}  (b) and (c), which is a characteristic of a Weyl semimetal.  The energy spectrum is shown in Fig. \ref{fig:weyl-en}.

Introducing a qubit with different coupling to each instance lattice, we obtain
\be
\mathscr{H}_e(k)=\delta \sigma_z
\ee
In Fig. \ref{fig:weyl-de}, $|\mathcal{L}_k(t)|^2$ is plotted for a case that clearly exhibits distinct features. There is a sharp valley when two band-touching points overlap, as shown in Fig. \ref{fig:weyl-de} (a); and there are two valleys when the band-touching points are momentum separated, as shown in Fig. \ref{fig:weyl-de} (b). For the other two cases shown in Figs. \ref{fig:weyl-en} (c) and (d), $|\mathcal{L}_k(t)|^2$ exhibits different features. A more detailed analysis is shown in Fig. \ref{fig:weyl-2d}. A common feature is that rapid oscillation in $|\mathcal{L}_k(t)|^2$ is accompanied with energy level crossing at the Fermi surface, whereas it tends to be one when the energy band is gapped. In addition, it is noted from Fig. \ref{fig:weyl-de} (d) that rapid oscillation around point $\Gamma$ corresponds to the degeneracy inside the bands. Moreover, this feature indicates that qubit decoherence can partially detect the degeneracy inside the energy band.

\section{conclusion}
We explicitly demonstrate the ability of qubit decoherence for detecting ground state degeneracy in topological systems. The primary observation is that $|\mathcal{L}_k(t)|^2$  exhibits distinct behaviors depending on whether or not energy level crossing occurs at the Fermi surface. This can be understood using a two band model, in which  $|\mathcal{L}_k(t)|^2$ depends strongly on $R(k)$, as shown in Eq.\eqref{dfk}. When the topology of the system varies because of the closing of the energy gap, corresponding to the vanishing of $R(k)$,   $\mathcal{L}(t)$ becomes singular. When the topology does not vary, $\mathcal{L}(t)$ tends to be one. In addition, $|\mathcal{L}_k(t)|^2$ can also reflect the degeneracy inside the band, as discussed previously, for the QWZ, KM, and Weyl models. Qubit decoherence can be used for indirect detection of degeneracy in many-body systems, for energy level crossing at the Fermi surface and for substantial band degeneracy.

$|\mathcal{L}_k(t)|^2$ was measured experimentally in an ultra-cold atomic ensemble \cite{lewenstein}. In recent years, several models have been proposed, in which the degenerate points were identified unambiguously by tunable spin-orbit couplings in optical lattices \cite{soc} or by measuring the variation in atom number \cite{ol, duca}. Fluctuation in atom number induces decoherence in a qubit, which can be measured using $|\mathcal{L}_k(t)|^2$.

In recent years, using the concepts and methods developed in quantum information, physicists have gained new viewpoints on exotic effects in condensed matter systems, as it is recognized that the wave function of many-body system is  more fundamental. This work is our contribution to this viewpoint. Future work involves determining whether there exists a similar method to detect topological numbers in condensed matter systems, \emph{e.g.}, Chern number. Different from the local feature of energy degeneracy, a global method must be developed to characterize the topological numbers. A possible method is to study decoherence of proper entanglement \cite{deentanglement}, as quantum entanglement is a manifestation of nonlocal features in quantum states.

The authors acknowledge the support of NSF of China (Grant Nos. 11005002, 11175032, 61475033) and NCET of Ministry of Education of China(Grant No. 11-0937).

\begin{widetext}

\begin{figure}[t]
\center
\includegraphics[width=8cm]{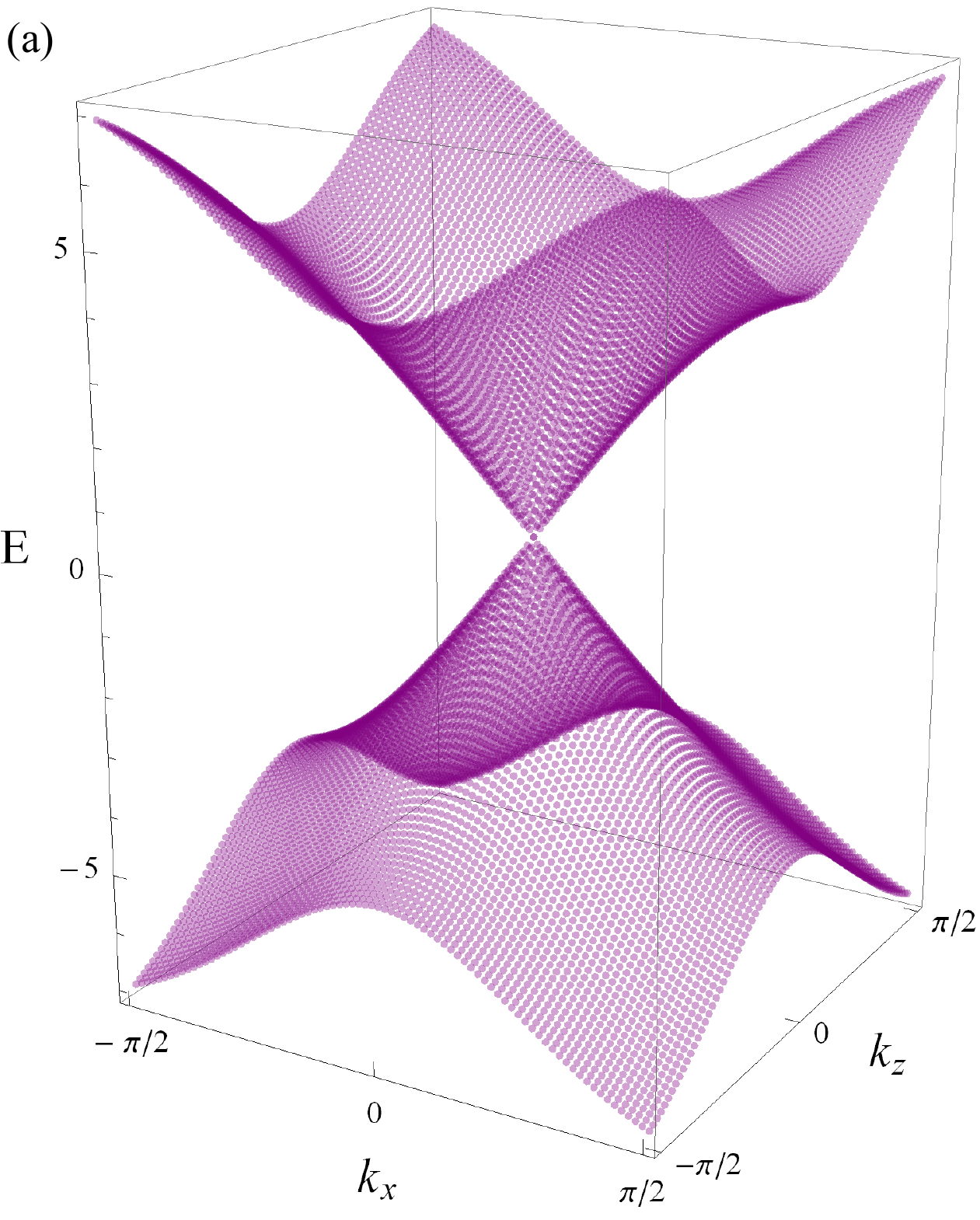}
\includegraphics[width=8cm]{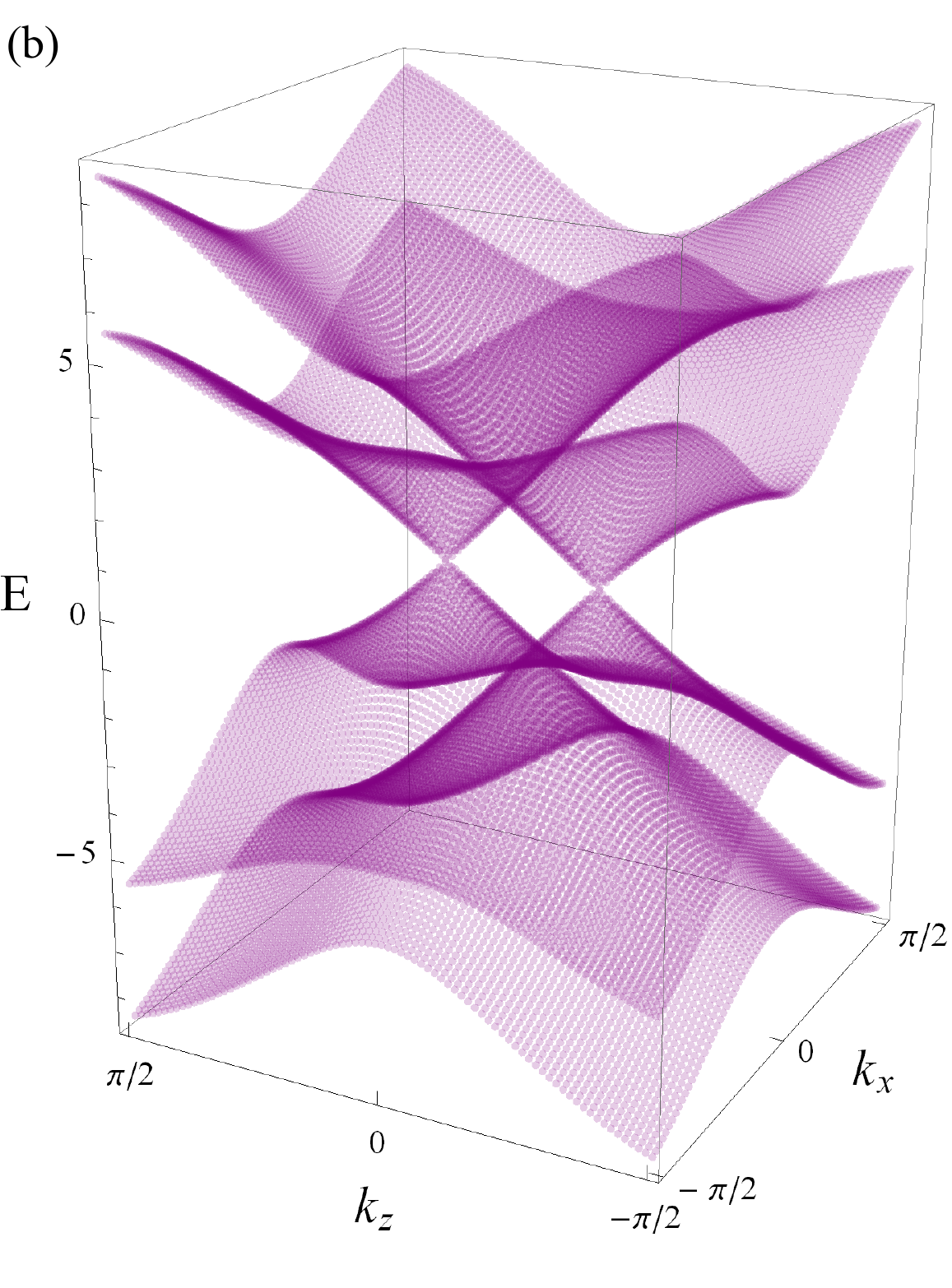}
\includegraphics[width=8cm]{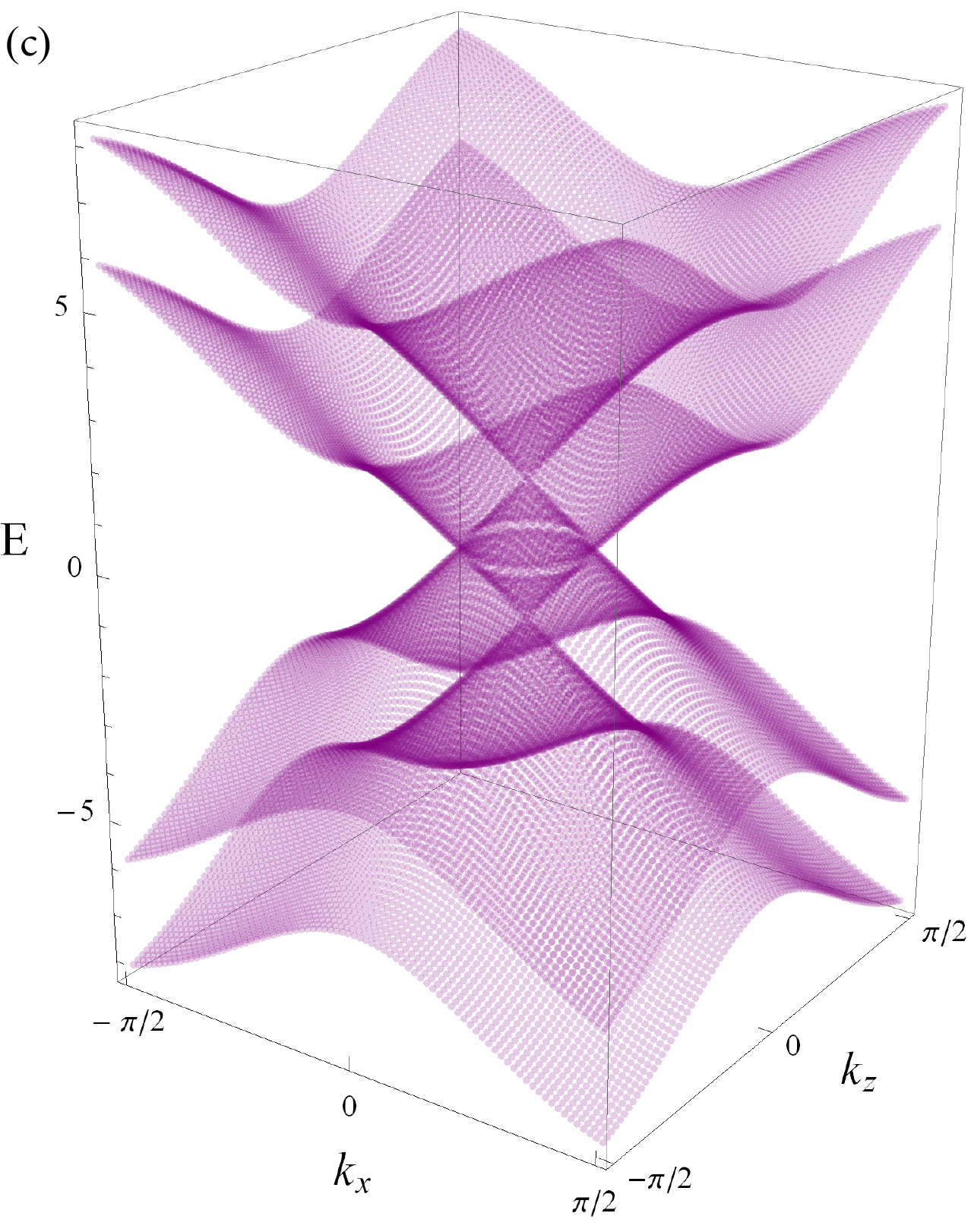}
\includegraphics[width=8cm]{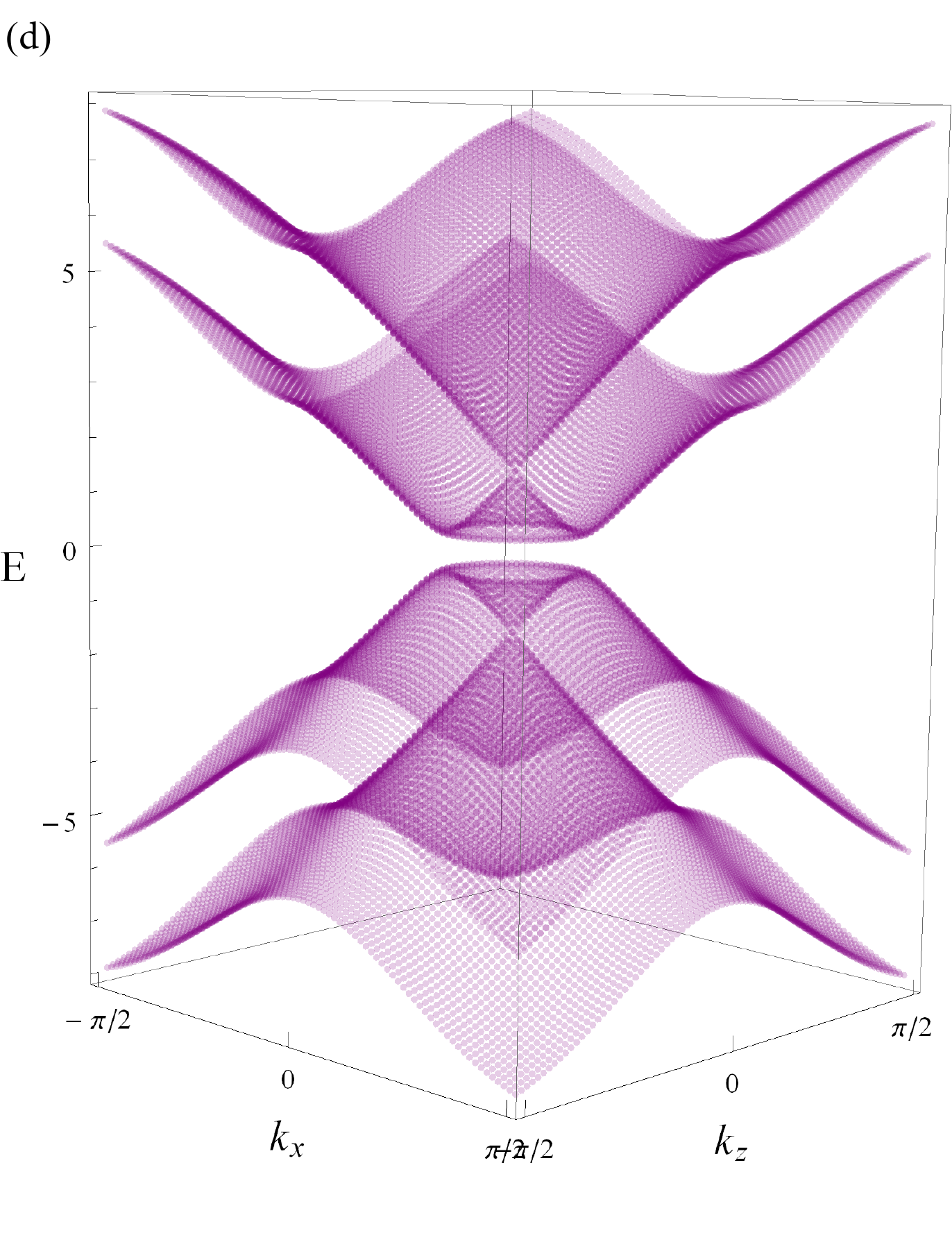}
\caption{(Color online) The energy spectrum of Eq. \eqref{weyl} for (a)  $\mathscr{H}_I(k)=0$ and $\varepsilon=6t$, in which the band-touching points are doubly degenerate; (b) $\varepsilon=6t$, $b_z=1.8t$ and $b_0=0$, in which the band-touching points are momentum-seperated; (c) $\varepsilon=6t$, $b_z=0$ and $b_0=1.4t$, in which the band-touching points are energy-seperated ; (d) $\varepsilon=5.5t$, $b_z=0$ and $b_0=1.4t$, in which an energy gap between upper and lower bands is opened.  For all plottings, $\lambda=\lambda_z=2$ and $b_1=b_2=0$ are chosen. The lattice constant is set to be 1.  }
\label{fig:weyl-en}
\end{figure}

\begin{figure}
\center
\includegraphics[width=8cm]{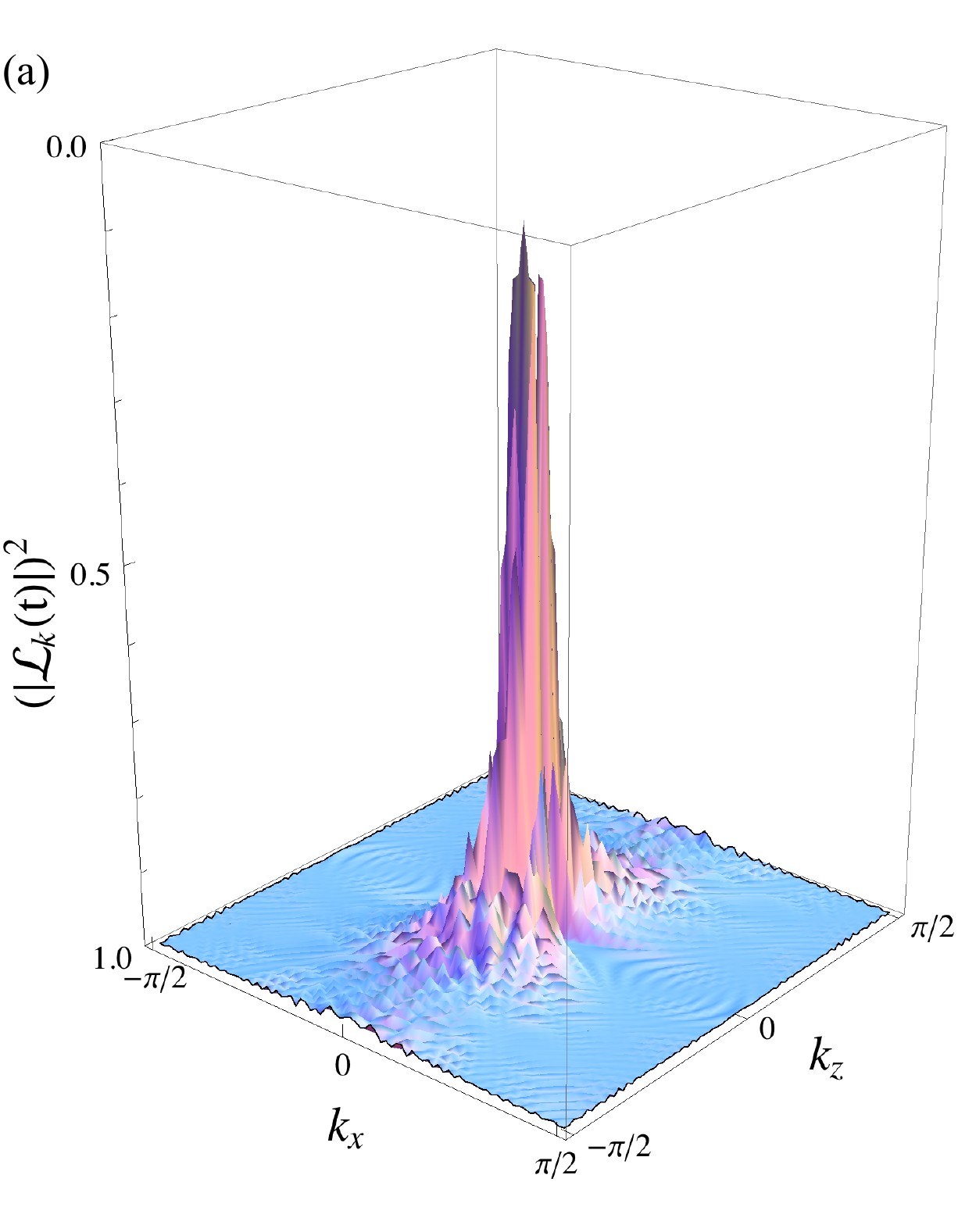}
\includegraphics[width=8cm]{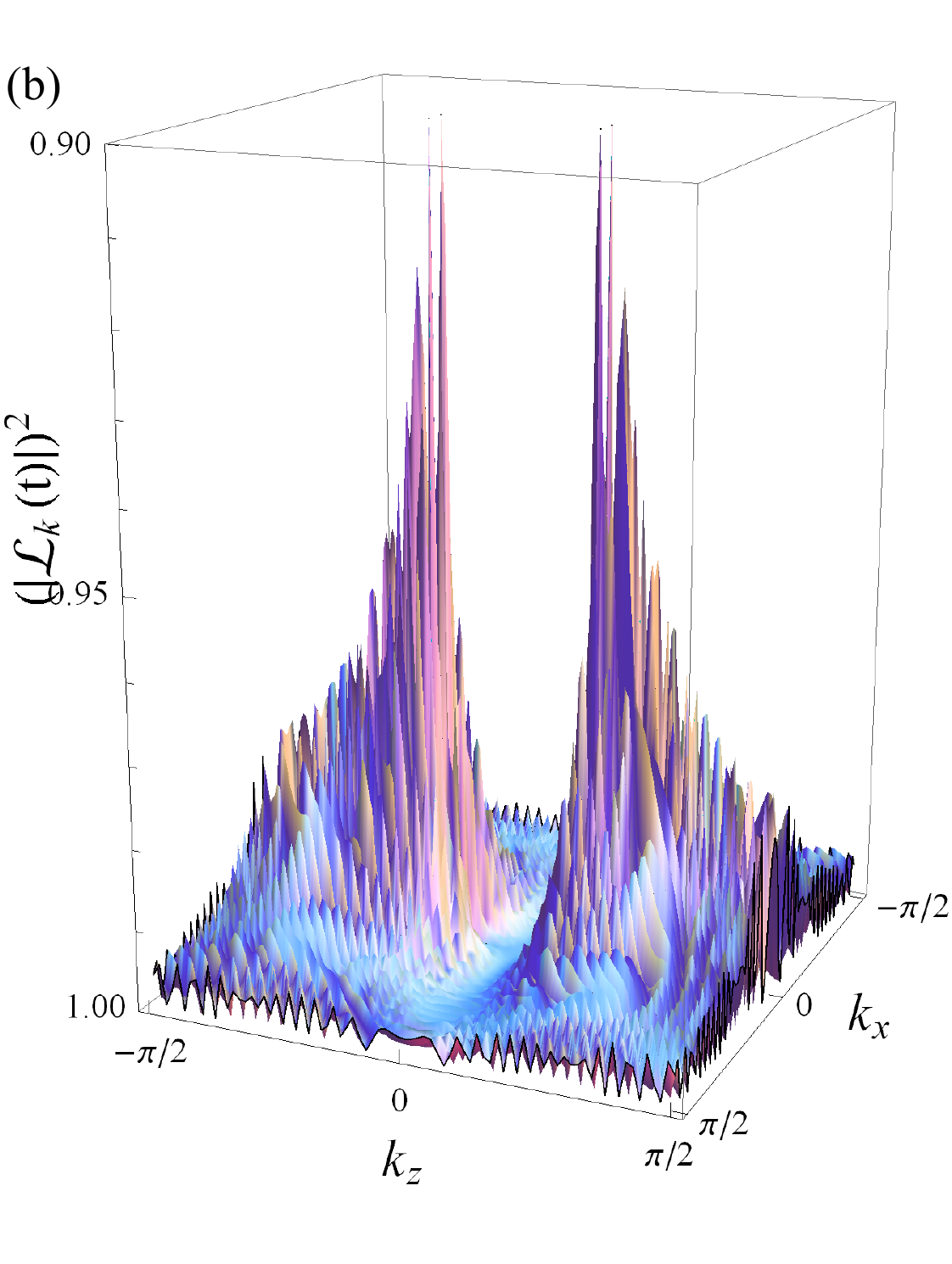}
\includegraphics[width=8cm]{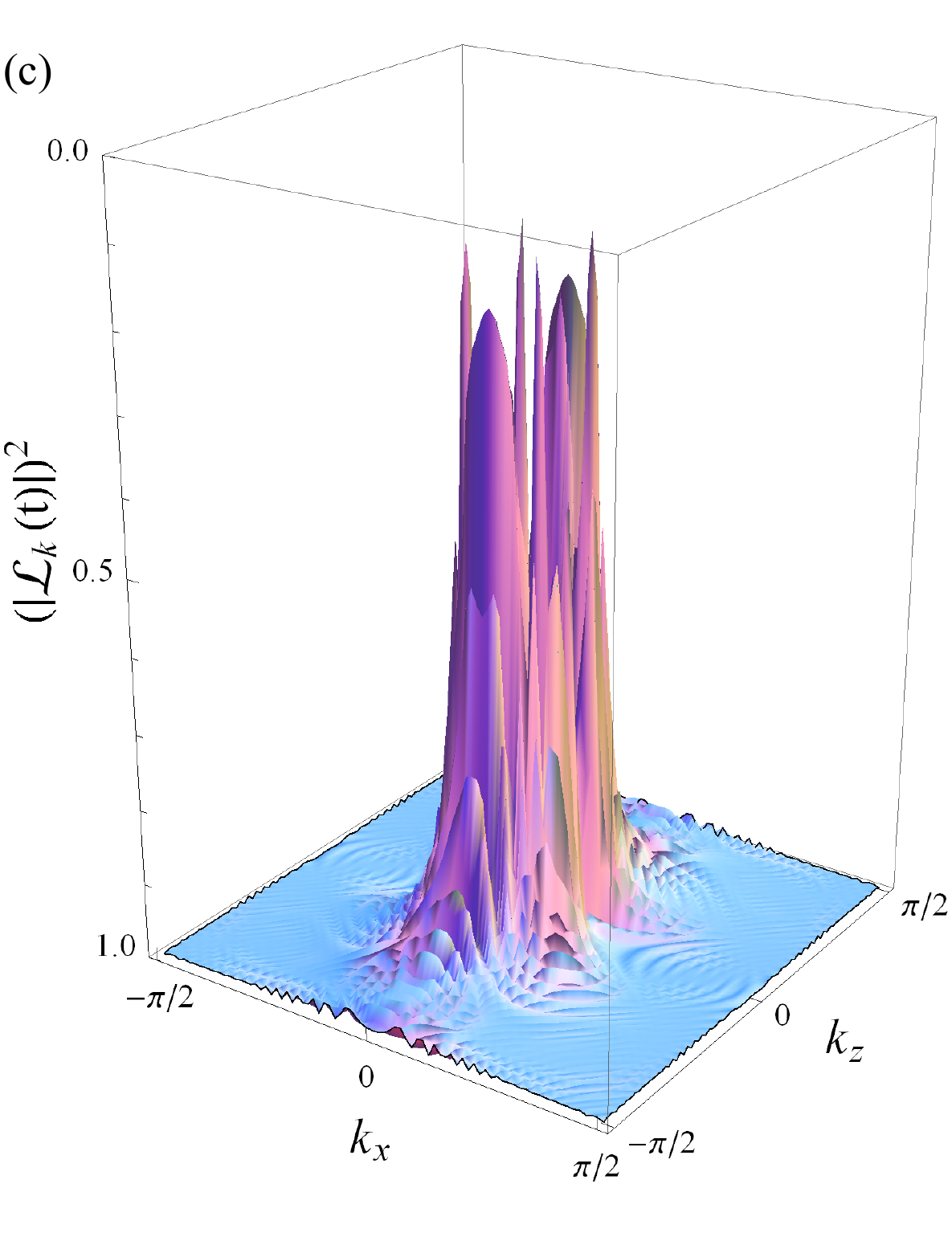}
\includegraphics[width=8cm]{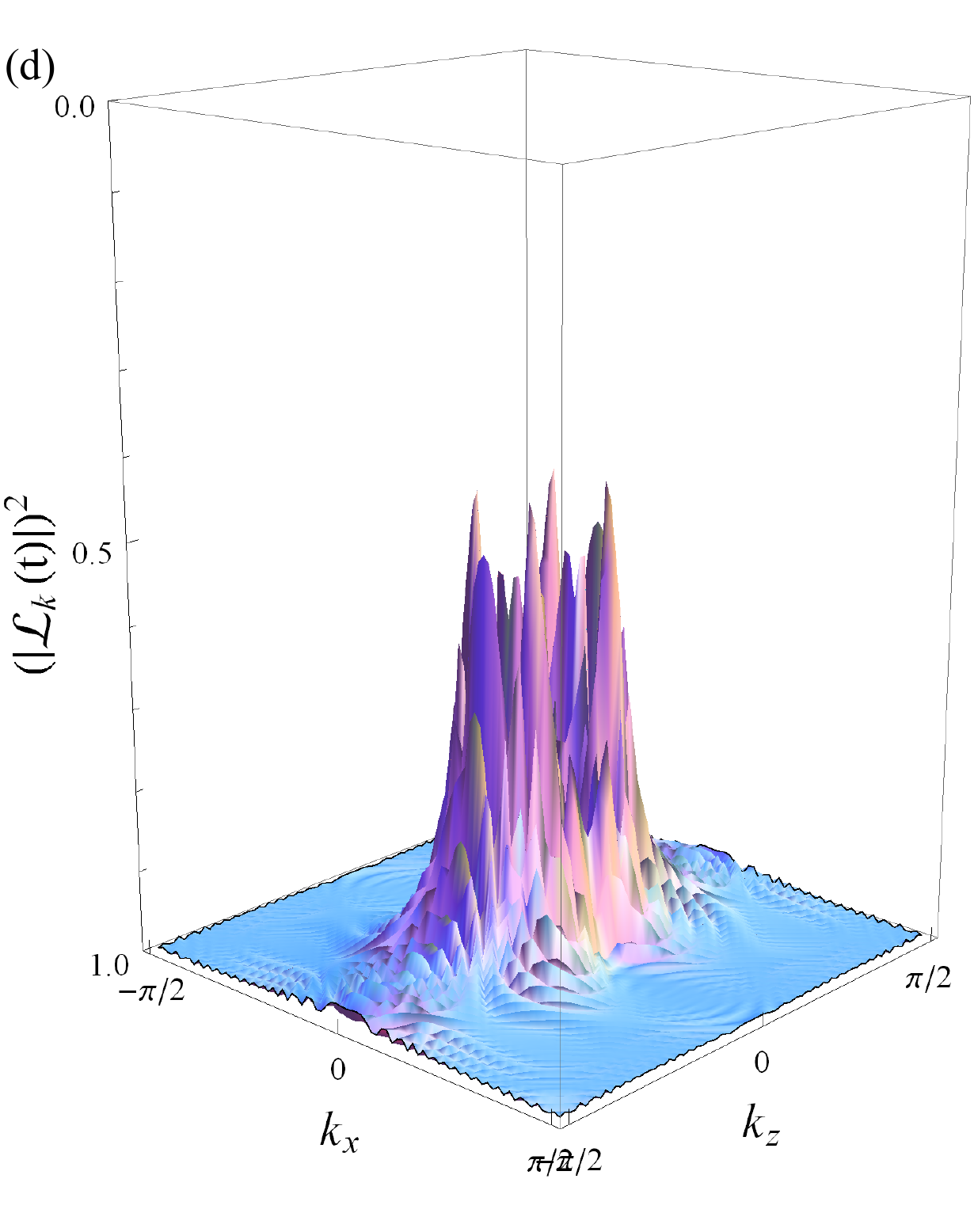}
\caption{(Color online) The decoherenc factor $|\mathcal{L}_k(t)|^2$ for Weyl semimetal with $\delta=0.5$. The parameters are same to that in plotting of Figs. \ref{fig:weyl-en}. Please note that the value of $|\mathcal{L}_k(t)|^2$ is shown by reversed order.}
\label{fig:weyl-de}
\end{figure}

\begin{figure}
\center
\includegraphics[width=8cm]{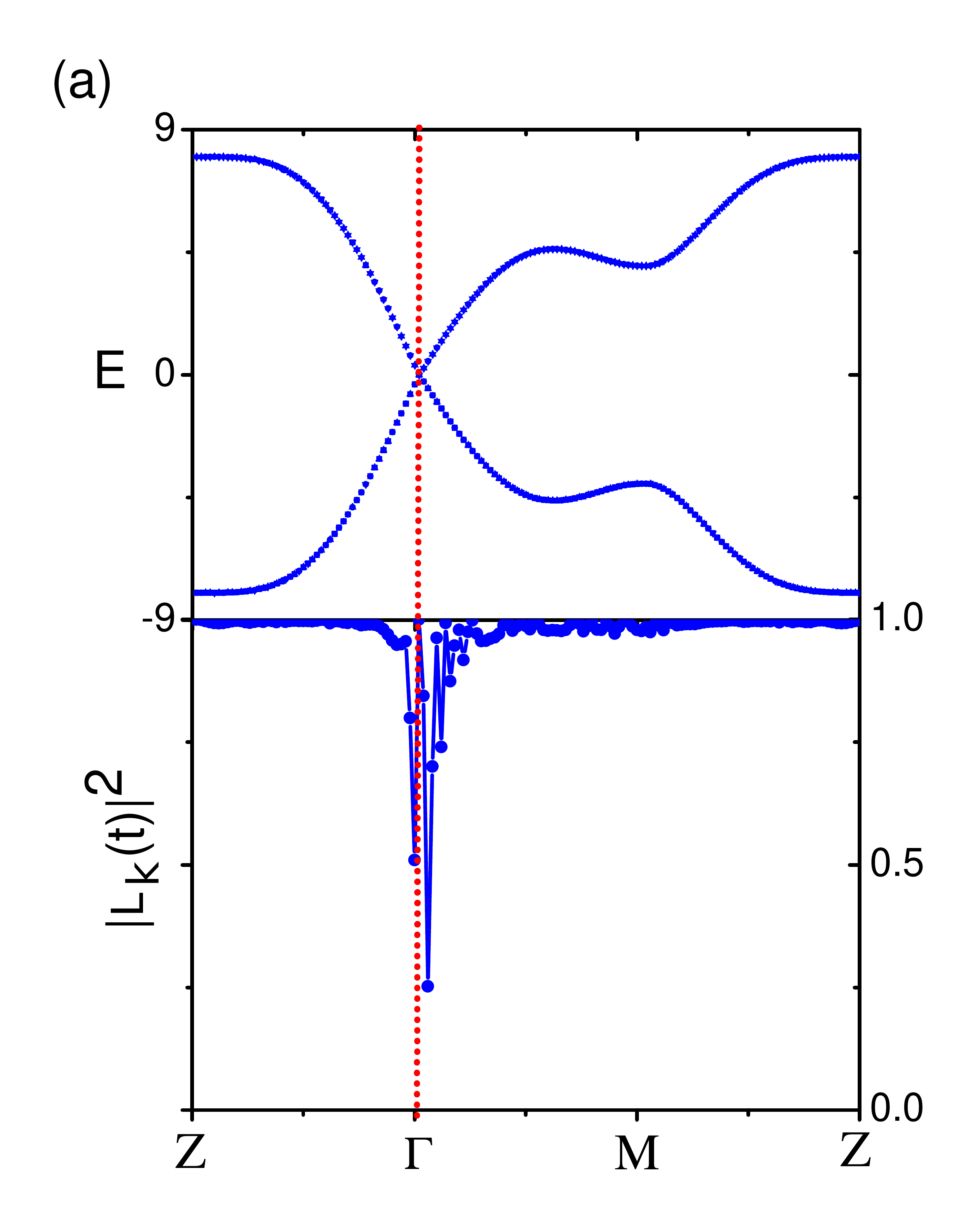}
\includegraphics[width=8cm]{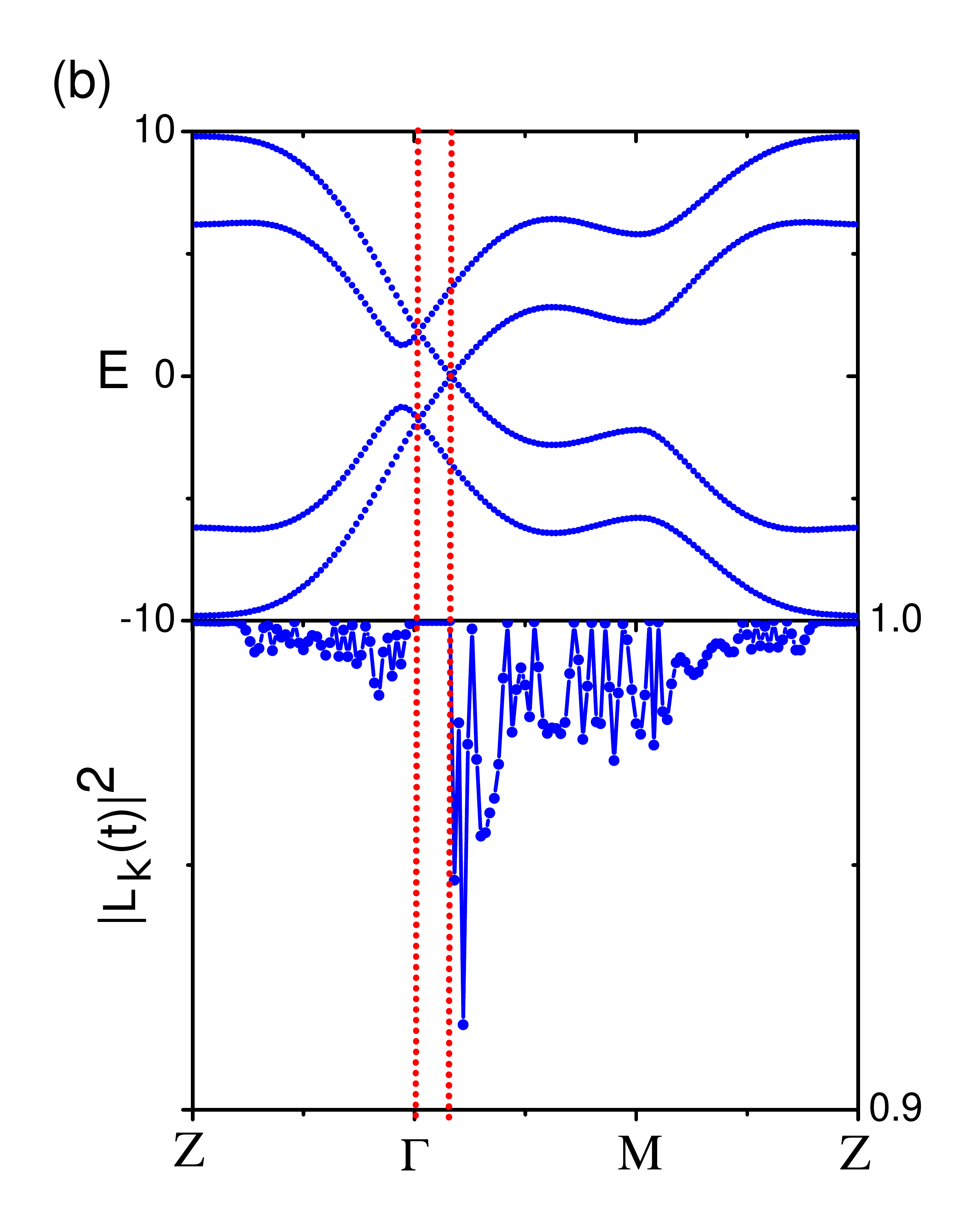}
\includegraphics[width=8cm]{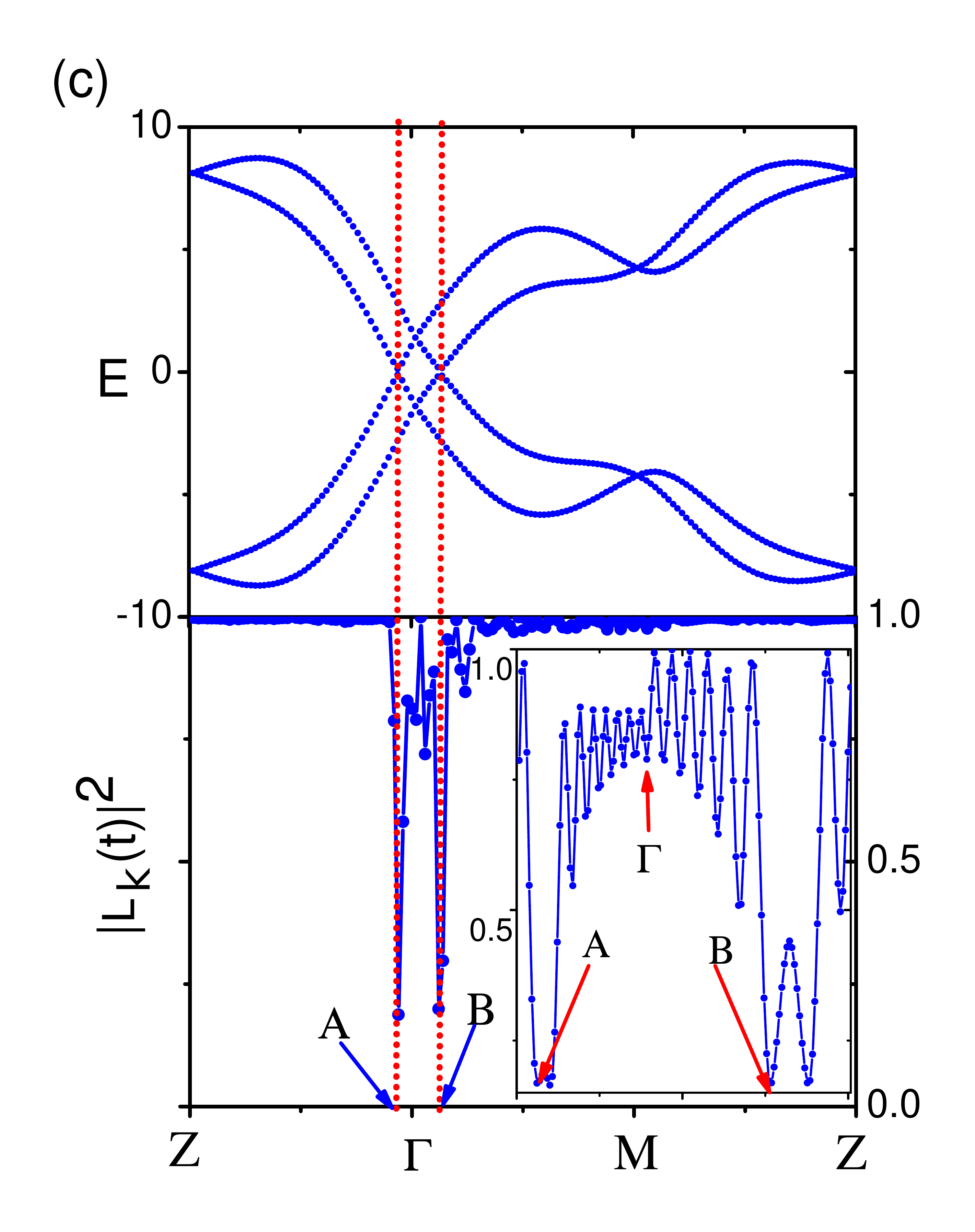}
\includegraphics[width=8cm]{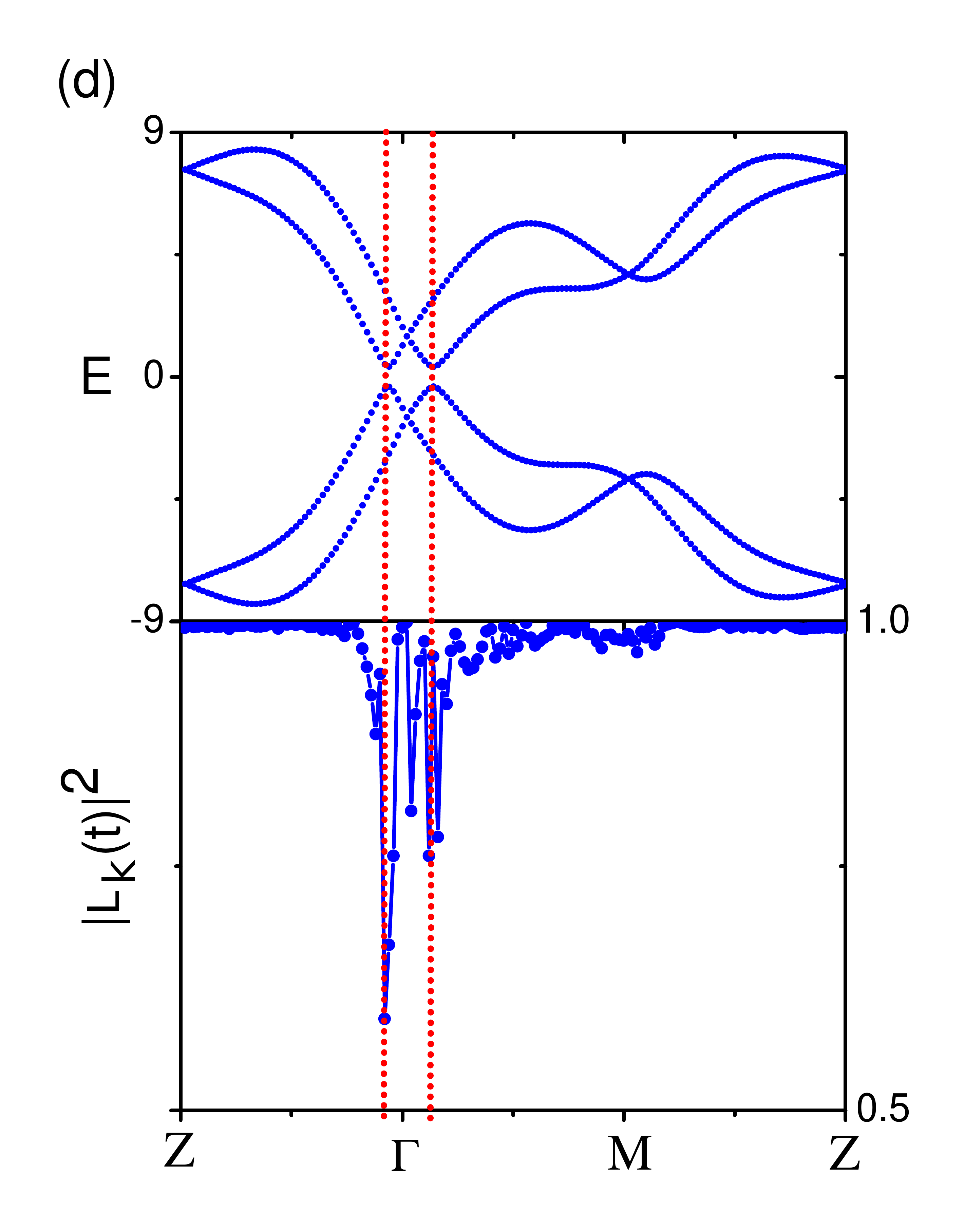}
\caption{(Color online) The direct comparison  between  $|\mathcal{L}_k(t)|^2$  and the energy band along the momentum path $(k_x, k_y, k_z)$: $Z=(\pi, 0, \pi)\rightarrow \Gamma=(0, 0, 0)  \rightarrow M=(0, 0, \pi) \rightarrow Z$. The parameters are same to that in Figs. \ref{fig:weyl-en} and \ref{fig:weyl-de}}.
\label{fig:weyl-2d}
\end{figure}

\end{widetext}

\end{document}